\documentclass[12pt]{article}
\usepackage{epsfig}
\usepackage{amssymb}

\def\gsim{\mathrel{\rlap {\raise.5ex\hbox{$ > $}}
{\lower.5ex\hbox{$\sim$}}}}
\def\lsim{\mathrel{\rlap {\raise.5ex\hbox{$ < $}}
{\lower.5ex\hbox{$\sim$}}}}

\newcommand{\be}{\begin{equation}}
\newcommand{\ee}{\end{equation}}
\newcommand{\bea}{\begin{eqnarray}}

\newcommand{\eea}{\end{eqnarray}}

\def\gappeq{\mathrel{\rlap {\raise.5ex\hbox{$>$}}
{\lower.5ex\hbox{$\sim$}}}}

\def\lappeq{\mathrel{\rlap{\raise.5ex\hbox{$<$}}
{\lower.5ex\hbox{$\sim$}}}}

\begin{document}

\begin{titlepage}

\begin{flushright}
%DFPD-00/TH/39 \\
%CERN-TH/00-?? \\
%CTP-TAMU-44/99 \\
%hep-ph/0008nnn
%eprint Draft-2\\
%25.03.2002
\end{flushright}

\vspace{0.1in}

\begin{center}
{\Large \bf  Prospects\\
 to measure neutrino oscillation
%curve
pattern\\
with  very large area underground detector\\
at very long baselines
%\footnote[1] {convener:~A. Zaitsev (IHEP,
%Protvino, Russia)

%~~E-mail address: zaitsev@mx.ihep.su}

%using  the JAERI-KEK  HIPA

}
\end{center}

\vspace{0.3in}

\begin{center}

 \vspace{.15in}
 {\large V. Ammosov, V. Garkusha, A. Ivanilov, V. Kabachenko,\\
  E. Melnikov, F. Novoskoltsev, A. Soldatov, A. Zaitsev\\}
%  $\,$and$\,$ G. Volkov$^{a,}$$^{b}$\\}
\vspace{.25in}
{ 
%$^{a}$

\sl Institute for High Energy Physics,\\
RU-142284  Protvino, Moscow region, Russia\\}
%{\it  $^b$ Universita di Padova, Dipartimento di Fisica Galileo Galilei,\\
%INFN, 35131 Padova, Italy\\}

 \vspace{2.0cm}

\end{center}

\begin{abstract}
The concept of a very long baseline neutrino experiment with quasi
monochromatic neutrino beam  and very large area underground
detector is discussed. The detector could be placed in the
existing 20~km tunnel at IHEP, Protvino. The High Intensity Proton
Accelerators (HIPA) which are planned to be built in Japan
(JAERI-KEK, baseline of $\sim 7000$~km) and Germany (GSI,
baseline of  $\sim 2000$~km) as well as the Main Injector at
Fermilab ($\sim 7600$~km) are considered as possible sources of
neutrino beams. The oscillations are analysed in the three-neutrino
scheme taking into account terrestrial matter effects. In the proposed 
experiment it is feasible to observe the oscillation pattern 
as an unique proof of the existence of neutrino oscillations. Precise 
measurements of disappearance oscillation parameters of the muon neutrinos 
and antineutrinos can be done within a reasonable time.

\end{abstract}

\end{titlepage}

\section{\bf  Introduction}
The observation of the atmospheric $\nu_{\mu}$
disappearance~\cite{SKA,MACRO} stimulated an
impressive number of proposals of new experiments with neutrino
beams from proton synchrotrons~\cite{UNK,K2K,MINOS,CNGS,JHF-SK,hipa} 
and from neutrino  factories~\cite{CNF}.
Main goals of this new generation of experiments are
the ultimate confirmation of neutrino oscillations and precise measurements
of oscillation parameters.

%It is obvious that only the experimental observation of  neutrino
%oscillation curve is real prove for the existence of
%the neutrino oscillations. Up to now  only disappearance of solar and
%atmospheric neutrinos is established which can be attributed to
%experimental biases and model dependence.

Very long baseline (VLBL) neutrino-oscillation experiments with
very large area (VLA) detectors are especially effective to
observe the neutrino oscillation patterns and to measure
precisely oscillation parameters. Here we consider the prospects to use
underground detector placed in the existing UNK tunnel  to carry
out these measurements.
%The detector will register muons from
%$\nu_{\mu}$ CC interactions in surrounding soil mainly. Expected
%effective mass will be at a $\sim 1~ Mt~ level$ for a few GeV
%neutrino energy.
As a source for neutrino beams we consider the
high intensity proton accelerators which are planned to be built
in Japan (JAERI-KEK, Tokaimura, baseline of $\sim 7000$~km) and
Germany (GSI, Darmstadt, baseline of $\sim 2000$~km) as well as
the existing Main Injector at Fermilab ($\sim 7600$~km).

In the second section the physics justification of the proposed approach
is given. The concept of a possible experimental lay-out including
a neutrino focusing system and UNK underground detector are described
in the third section. Physics performance is outlined in the fourth
section. The last section is devoted to conclusion.

\section{\bf  Physics motivation}

We propose to measure the $\nu_{\mu}$ and $\bar \nu_{\mu}$
survival probabilities $P(\nu_{\mu}(\bar
\nu_{\mu})\rightarrow ~\nu_{\mu}(\bar \nu_{\mu}))$ as a function
of the neutrino energy. The UNK underground detector will register
muons from $\nu_{\mu}$ and $\bar \nu_{\mu}$ charged current (CC) interactions
in surrounding soil mainly. Expected effective mass will be at a
$\sim 1$~MTon level for a few GeV neutrino energy.

The motivation of the oscillation approach is based on the observation of
atmospheric $\nu_{\mu}$ disappearance and a small level, if any, of
reactor $\nu_{e}$ disappearance.
Existing experimental data for the muon (electron) neutrino disappearance are 
analyzed in  terms of the simple formula for oscillation probability 
$P(\nu_\mu(\nu_e)\to\nu_\mu(\nu_e))=1-I_{\mu (e)}\cdot \sin^2(1.27\cdot \Delta
m^2\cdot L/E)$ between two neutrino of different flavours (see Table~1 for the
experimental restrictions on the parameters).

\begin{table}[th]
\begin{center}
{\bf Table 1.} Summary of experimental results on 
neutrino disappearance.\\[1ex]
\begin{tabular}{|c|c|c|}
\hline
Experiments          & Best values & Allowed region\\\hline
disappearance of     & $I_{\mu}=1.$ & $I_{\mu}>0.84$~(99\%~CL) \\
atmospheric $\nu_{\mu}$~\cite{SK}&$\Delta m^2=2.5\cdot 10^{-3}$~eV$^2$ &
 $1.2\cdot 10^{-3}<\Delta m^2<4\cdot
 10^{-3}$~eV$^2$~(99\%~CL)\\\hline
disappearance of     & --   & $I_{e}<0.1$~(90\%~CL) \\
reactor $\bar \nu_{e}$~\cite{chooze}&        &   \\
\hline
  \end{tabular}
 \end{center}
\end{table}

At $\Delta m^2=2.5\cdot 10^{-3}$~eV$^2$ and with a few GeV neutrino energy the
oscillation length is in the region of $10^3 - 10^4$~km.
Therefore, the
measurement of the $P(\nu_{\mu}(\bar \nu_{\mu})\rightarrow
~\nu_{\mu}(\bar \nu_{\mu}))$ probability as a function of neutrino energy
is a natural way to observe the neutrino oscillation pattern and to measure
precisely the oscillation parameters for this pattern in the VLBL neutrino
experiments.

The observation of
atmospheric $\nu_{\mu}$ disappearance is just an indication on
neutrino oscillations but not a definite proof of it.
Within the proposed approach different aspects of neutrino physics
could be studied:
\begin{itemize}
 \item Measurement of $\nu_{\mu}$ and $\bar \nu_{\mu}$
 disappearance patterns.

In the case of a simple oscillation pattern
$P(\nu_{\mu}\rightarrow
~\nu_{\mu})=1-I_{\mu}\cdot
\sin^2(1.27\cdot \Delta m^2\cdot L/E) $
 the precise measurements of $\Delta m^2$ and $I_{\mu}$ can be done for
$\nu_{\mu}$ and $\bar \nu_{\mu}$.
\item Search for no oscillations or non-standard oscillations.

The models of neutrino disappearance by decay~\cite{nude_barger}
or nonstandard oscillations due to flavour changing neutrino
interactions~\cite{Gago} could be checked with high sensitivity.
Observation of more complicated oscillation pattern could be a
signal, for example, of large extra dimensions~\cite{dvali}.
\item Measurement of the difference $\Delta_\mu=
P(\nu_{\mu}\to\nu_{\mu})-P(\bar \nu_{\mu}\to\bar \nu_{\mu})$.

It can come from a fake $CPT$ violation due to the terrestrial matter 
effects~\cite{xing} and can be from a genuine $CPT$ violation if it 
exists~\cite{CPT}. As fake $CPT$ violation depends on the CP-violating phase, 
the  proposed experiment could provide information on it as well.
\end{itemize}

%It is expected now as it is seen from the Table~\ref{tab1} that
%\begin{itemize}
%\item $\nu_{\mu}$  mixes $\sim 100\%$ with $\nu_{\tau}$
%and only $ < 5\%$ with $\nu_{e}$;
%\item $\nu_{e}$  does not like to mix nor with $\nu_{\mu}~(< 5\%)$,
%nor with $\nu_{\tau}~(< 5\%)$.
%\end{itemize}

Below we consider the simple oscillation model in vacuum as the
baseline, concentrating on the ability to observe the oscillation
pattern.

Fig.~1 shows the $\nu_{\mu}$ and  $\nu_{e}$ survival probabilities as
functions of $L/E$. Arrows indicate the maximal $L/E$ values
which can be reached with the MINOS~\cite{MINOS}, the CERN-Gran
Sasso (CNGS)~\cite{CNGS}, the JHF-SK~\cite{JHF-SK} and the CERN
Neutrino Factory (CNF)~\cite{CNF}
planned experiments. \\
\centerline{
\begin{tabular}{c}
 \epsfig{file=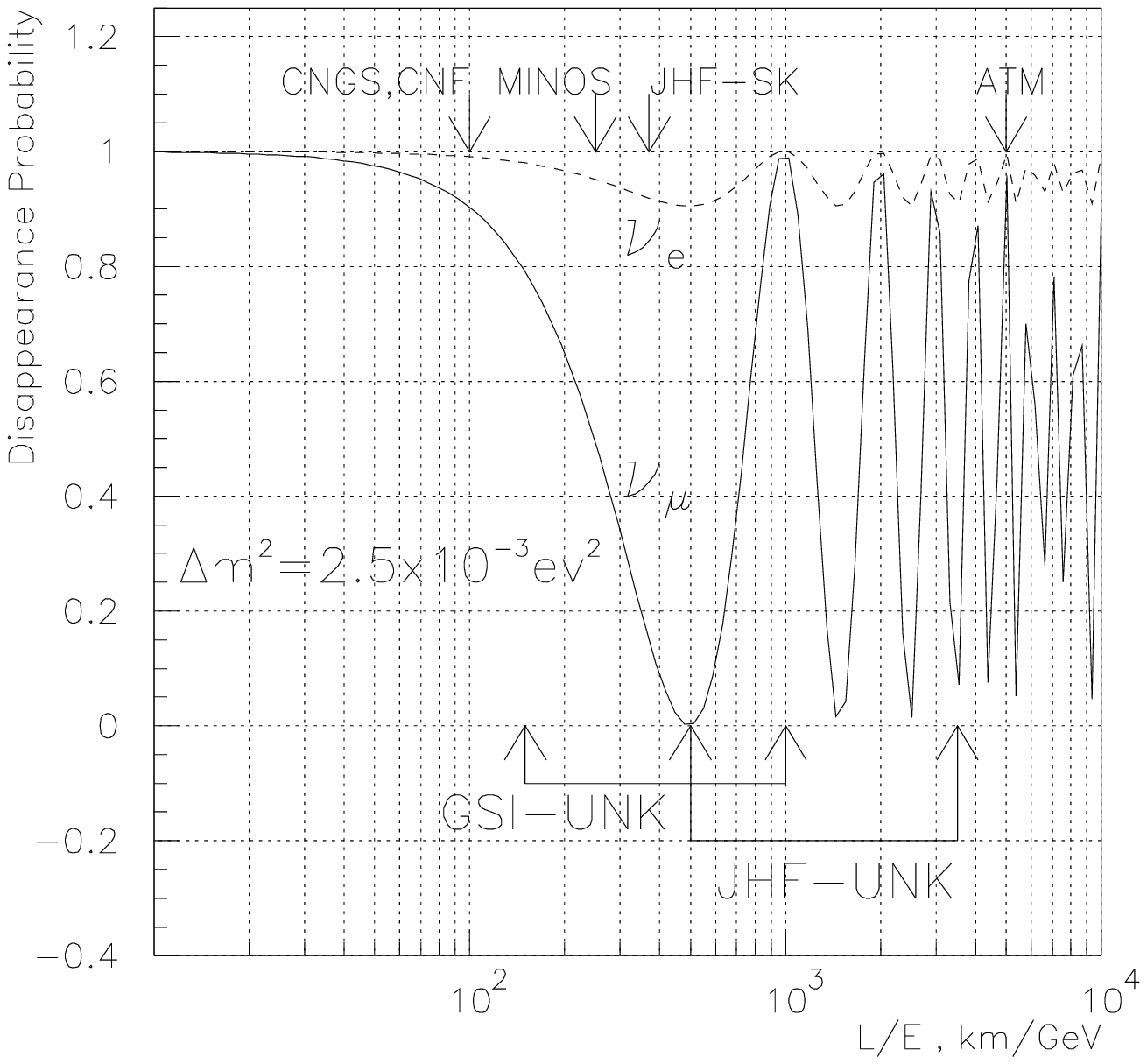,width=9.3cm} \\
{\bf Fig. 1.} $P(\nu_\mu\to\nu_\mu)$ and $P(\nu_e\to\nu_e)$ survival 
probabilities.
\end{tabular} }
\vspace{0.3cm}

As it is seen in the figure CNF~\cite{CNF},  CNGS~\cite{CNGS} and
MINOS~\cite{MINOS}
%all of these
projects
%excluding the JHF-SK
are at the beginning of the first wave of the oscillation curve, where 
oscillation patterns are not seen. The JHF-SK project~\cite{JHF-SK} has some 
chance to observe an oscillation pattern for 
$\Delta m^2>3\cdot 10^{-3}$~eV$^2$.
%Therefore expected "disappearance"
%or "appearance" of those accelerator projects can not convince
%everybody with existence of neutrino
%may be explained by some biases.
%For example, wrong pointing of
%neutrino beam to a far detector.

As one can see in the figure the oscillation pattern in principle
could be measured for the atmospheric $\nu_{\mu}$. However to do this,
a precise reconstruction of the neutrino direction is needed to
obtain the neutrino pass length. In
the SK and MACRO detectors the neutrino direction is defined
from the direction of muons which have wide angular distribution at low
neutrino energy ($\langle E_{{\nu}_{\mu}}\rangle=2.4$~GeV). Therefore, only
the average disappearance was observed.
% for upward-going neutrinos.
   Thus the oscillation curve is clearly visible if
\begin{itemize}
\item oscillation phase $\phi_{osc} =1.27\cdot \Delta m^2\cdot
L/E\approx (2n+1)\cdot \pi/2,~n=0,1,....$;
\item  error $\delta \phi_{osc} < \pi/4$.
\end{itemize}
Translation of these conditions to the energy resolution gives
$${\delta E_{\nu}\over E_{\nu}} < {1\over 2(2n+1)}.$$
It means that the soft restriction is for $n=0$ and 1,
i.e., for the first and the second minima.

If we look at the Earth globes, the 1-st condition can be fulfilled
if neutrino beams are sent to Protvino from the Fermilab MI, the
JHF or the GSI (see expected JHF-UNK and GSI-UNK $L/E$ regions on
Fig.~1). In the Table~2
 the possible characteristics of
beams from the MI, the JHF and the GSI are presented.

\begin{table}[th]
 \begin{center}\begin{minipage}{\textwidth}
 {\bf Table 2.} Comparison of neutrino sources for
  UNK VLBL neutrino oscillation experiment.
  \end{minipage}\\[1ex]
  \begin{tabular}{|l|c|c|c|}
  \hline
 Parameter            & JHF, Japan & GSI, Germany$^{*}$&MI FNAL, USA\\\hline
 Baseline from the UNK, km& 7000       &2000         & 7600\\\hline
 Tilt angle, degree& 33         &9            & 34\\\hline
 Proton momentum, GeV$/c$& 50         &50           & 120\\\hline
 Cycle time, s      & 3          &3            & 2\\\hline
 One turn time, $\mu$s& 5        &4            & 10\\\hline
 Proton intensity/spill&$3.3\cdot10^{14}$&$1\cdot10^{14}$&$4\cdot10^{13}$
                                                      \\\hline
 Protons/s&$1\cdot10^{14}$&$3\cdot10^{13}$&$2\cdot10^{13}$\\\hline
 $\pi^+$ yield/$p$ at 7 GeV$/c$& 0.05&0.05       & 0.12\\\hline
 $\pi^+$ yield/s at 7 GeV$/c$&$5\cdot10^{12}$&$1.7\cdot10^{12}$
 &$2.4\cdot10^{12}$
                                                      \\\hline
 duty factor & $1.7\cdot10^{-6}$ &$1.3\cdot10^{-6}$&$5\cdot10^{-6}$\\\hline
  \end{tabular}
\begin{flushleft}
%\\
 $^{*}$ assumed parameters
\end{flushleft}
 \end{center}
\end{table}

%The neutrino event rate is proportional to $(\pi~ yield/sec)/L^2$
%and the rate/background ratio is proportional to
%$rate/(duty~factor)$. As it is seen from the Table the GSI and the
%JHF cases are in favour for these values.
%the signal/noise ratio is
%is 6-7 times
% better for the JHF  case due to high proton intensity and small duty factor.
%In case of the GSI the lower proton intensity is overcompensated by shorter
%distance between the UNK detector and the GSI.
Below we consider the JHF and the GSI as  neutrino beam
sites in comparison.
% for definiteness. For the GSI
%the numbers of events can be
%scaled easily.
%To satisfy the 2-nd condition  different energy settings of the
%neutrino beam should be done. As it has been shown in section 3.1
%it is possible for the Narrow Band Beam.
% Figures 2-4 show ideal nm
%disappearance curves and also smeared curves for WBB (så/å=0.36)
%and NBB (så/å=0.15) as a function of neutrino energy for
%Dm2=0.5þ10-3, Dm2=2.2þ10-3 and Dm2=6þ10-3 ev2, respectively. For
%human spirit the NBB curves are more convincing even with ~3 times
%lower statistics.
%   Crosses show possible NBB settings for oscillation curve measurement.
\newpage
\section{\bf  Experimental lay-out}
\subsection{\bf  Concept of experiment}
Sketch of the general experimental lay-out is shown in Fig.~2.\\

\centerline{ 
\begin{tabular}{c}
\epsfig{file=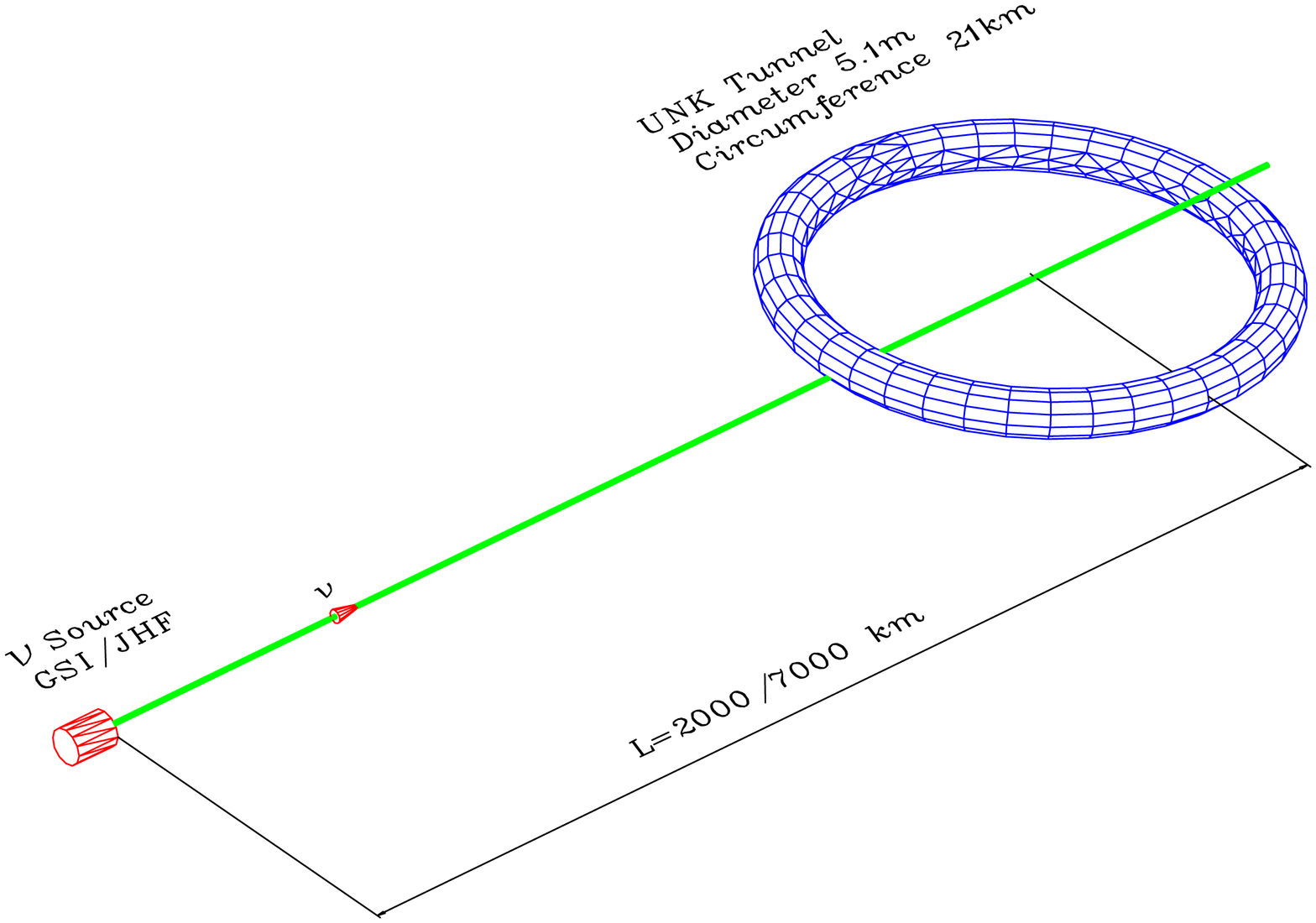,width=7.0cm,angle=90} \\
 {\bf Fig. 2.} General experimental lay-out.
 \end{tabular}
}
\vspace{0.3cm}
 Far neutrino source from the JHF (baseline of $7000$~km) or
GSI (baseline of $2000$~km) have to point to the UNK tunnel which is
$\sim 50$~m underground
 in average. For such kind of distances, to have enough statistics,
we should think about a detector
with effective mass at the 1~MTon  level.
 In such a case it is possible to use  the surrounding
soil of the UNK tunnel as neutrino target. It is proposed
to cover the UNK tunnel walls by scintillation counters.
For such an experimental concept, the UNK detector, which can
be considered as a huge double scintillator counter, will count mainly
muons from $\nu_{\mu}$ ($\bar \nu_{\mu}$) CC interactions in
the surrounding soil.

The circumference length of the UNK tunnel is $\sim 21$~km,
and the diameter is $\sim 5$~m.
It means that an area of about  $\sim 10^5$~m$^2$ can be used
for a very long baseline neutrino experiment. Dimensions
and time resolution of individual scintillation counters
will allow us
to select muons from the expected neutrino direction.

To suppress the cosmic ray background the fast (one turn)
extraction of the proton beam and the synchronization of a
neutrino source and of the UNK detector~\cite{GPS}
%using the GPS~\cite{GPS}
 are ultimately needed.

To observe the neutrino oscillation pattern we intend to use
a few (at least three) neutrino energy settings with
a reasonable energy resolution and just count the double
coincidences in the UNK detector. Ratios of the measured to expected
counts at different energy settings will provide
the possibility to observe $\nu_{\mu}$ and $\bar \nu_{\mu}$
oscillation patterns in the disappearance mode.

If we assume that at least the $n=1$ minimum should be detected
then
$${\delta E_\nu\over E_\nu} < 0.15.$$
The ideal $\nu_{\mu}$
disappearance curves and also the smeared curves for
$\delta E_{\nu}/E_{\nu} = 0.36$ (a WB like beam)
and  $\delta E_{\nu}/E_{\nu} = 0.15$ (a NB like beam)
as a function of the neutrino energy for
$\Delta m^2=2.5\cdot 10^{-3}$~eV$^2$
for the GSI and JHF cases are shown in Fig.~3.\\

\centerline { 
\begin{tabular}{cc}
\epsfig{file=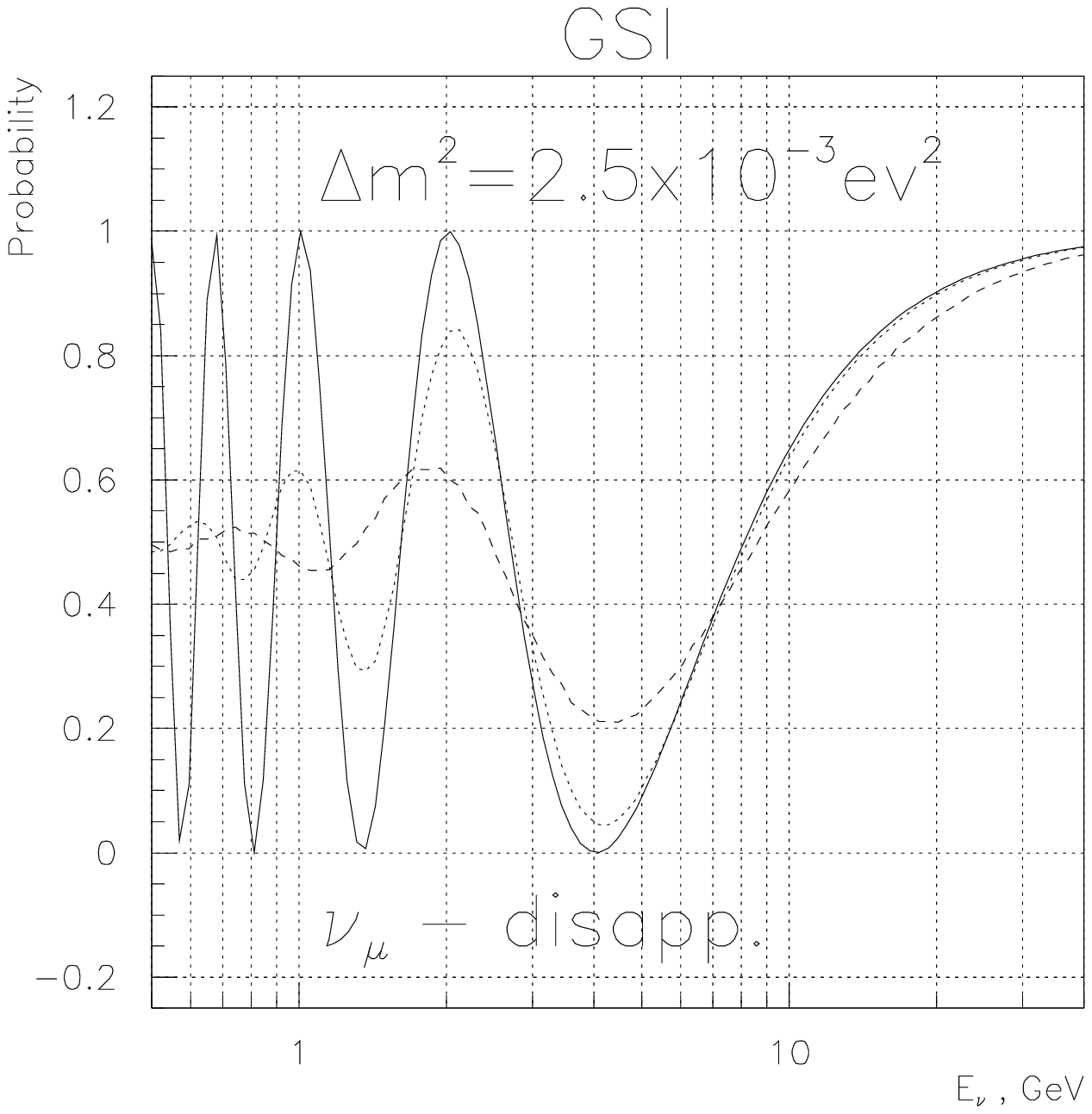,width=0.5\textwidth} &
\epsfig{file=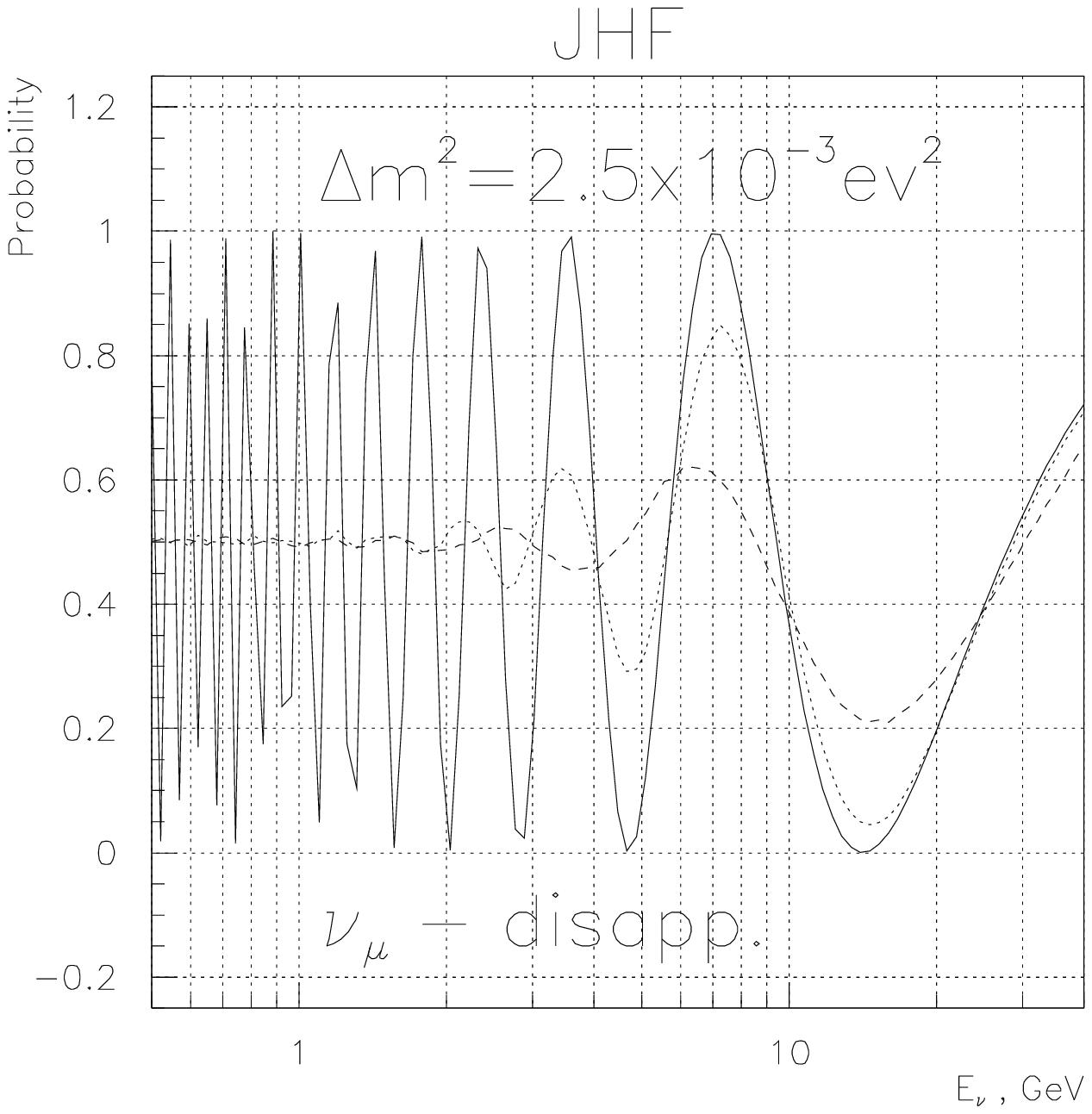,width=0.5\textwidth}\\ 
\multicolumn{2}{c}{
\begin{minipage}{\textwidth}
{\bf Fig.~3.} Ideal (solid line), smeared for $\delta E_{\nu}/E_{\nu} = 0.36$ 
(dashed line) and for $\delta E_{\nu}/E_{\nu} = 0.15$ (dotted line) 
oscillation curves for the GSI and the JHF cases.
\end{minipage} }
\end{tabular} }
\vspace{0.3cm}

 The importance of the $\delta E_{\nu}/E_{\nu} = 0.15$ condition
for a clear observation of the neutrino oscillation pattern is seen from this 
figure.

The proposed concept can be considered as
a further development of the atmospheric neutrino experiments with
upward-down-going muons. Here we already know:
\begin{itemize}
\item the exact distance from source to detector;
\item the type of neutrino ($\nu_{\mu}$ or $\bar \nu_{\mu}$);
\item the neutrino energy.
\end{itemize}

%Dm2=0.5þ10-3, Dm2=2.2þ10-3 and Dm2=6þ10-3 ev2, respectively. For
%human spirit the NBB curves are more convincing even with ~3 times
%lower statistics.

%    Fig.? shows the transverse cross-section of the UNK tunnel
%A detector should be rather course. It is intended to use the TOF
%and energy deposition in scintillator for event identification.

\subsection{\bf  Concept of the neutrino focusing system}
To estimate the possible $\nu_\mu$ and $\bar \nu_\mu$ event rates in the UNK 
detector we used the focusing system based on the magnetic horns, which were
optimized for the NuMI Project~\cite{our-preprint}. This focusing system 
(Fig.~4) can provide desired $\delta E_{\nu}/E_{\nu} = 0.15$ condition.

%To keep the $\delta E_{\nu}/E_{\nu} = 0.15$ condition the focusing system
%%, named "Dog Leg",
%based on the magnetic horns, which were
%optimized for the NuMI Project~\cite{our-preprint}, was used to
%estimate the possible $\nu_\mu$ and $\bar \nu_\mu$ event rates in
%the UNK detector (Fig.~4).

%Here we consider the JHF-UNK case for definiteness.
The 50~GeV primary proton beam
%with intensity of $10^{21}$ protons per year
is used to produce pions in the $\sim 2$ interaction lengths
graphite target T. Two 3~m long magnetic horns H1--H2 with
parabolic shaped inner conductors are used to focus the resulting
pion beam down a drift space where pions will decay to muon
neutrinos. Four 2~m long dipoles B1--B2 with 400~mm gaps are used
to obtain a beam of secondaries with relatively small $\Delta p/p$. The total 
length of the decay region is equal to 400~m
including the 40~m length target area and the 360~m length 
decay pipe of 1~m radius.

This focusing system may be tuned to different
neutrino energy ranges by scaling the dipole currents and
by corresponding adjustment of the target location, keeping the
current in both horns at its nominal value of 200 kA. Energy
spectra of $\nu_\mu$ and $\bar \nu_\mu$
CC events in the UNK detector are shown in Fig.~5 for
a few different neutrino energy settings.

\centerline{
\begin{tabular}{c}
\epsfig{file=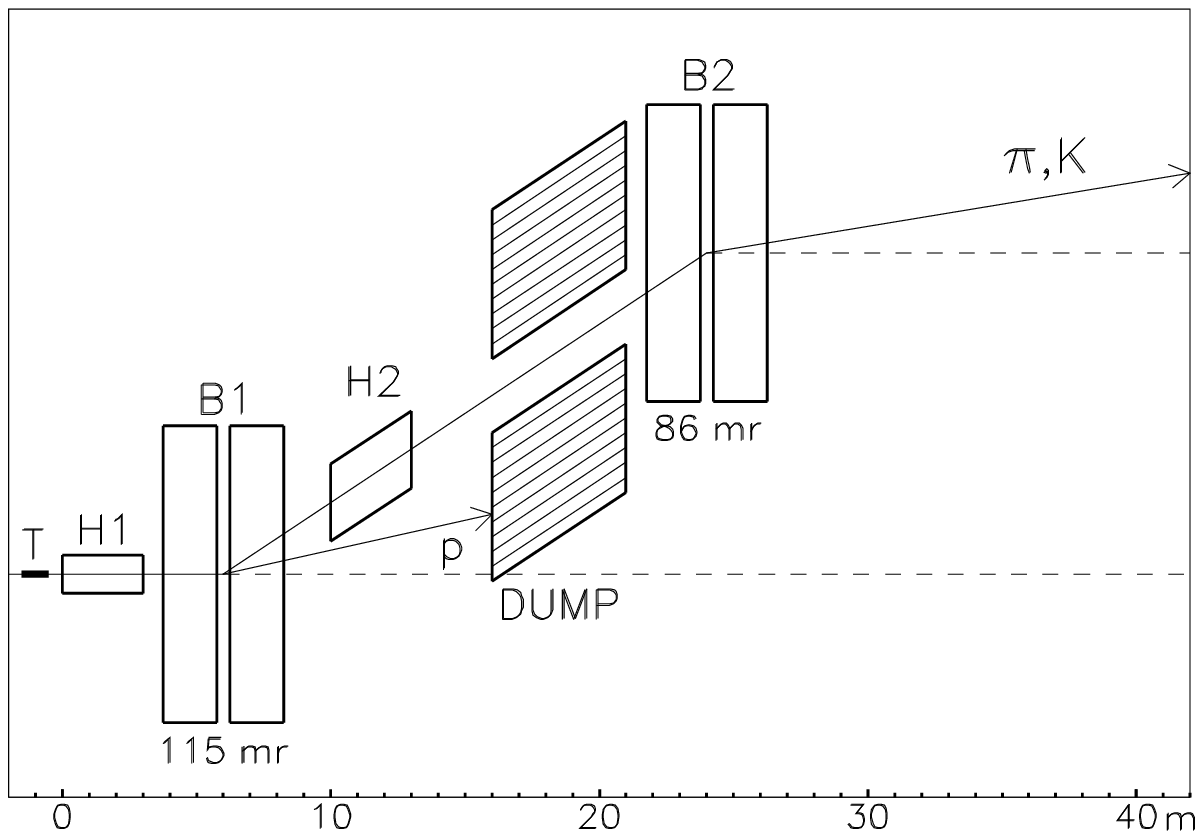,width=10.0cm} \\
{\bf Fig. 4.} The layout of the focusing system.
\end{tabular} }
\vspace{0.3cm}
%Due to uncertainties in
%hadron production cross sections, a systematic error of order 20\%
%should be applied to the $\nu_\mu$ rates.

\centerline{
\begin{tabular}{c}
\epsfig{file=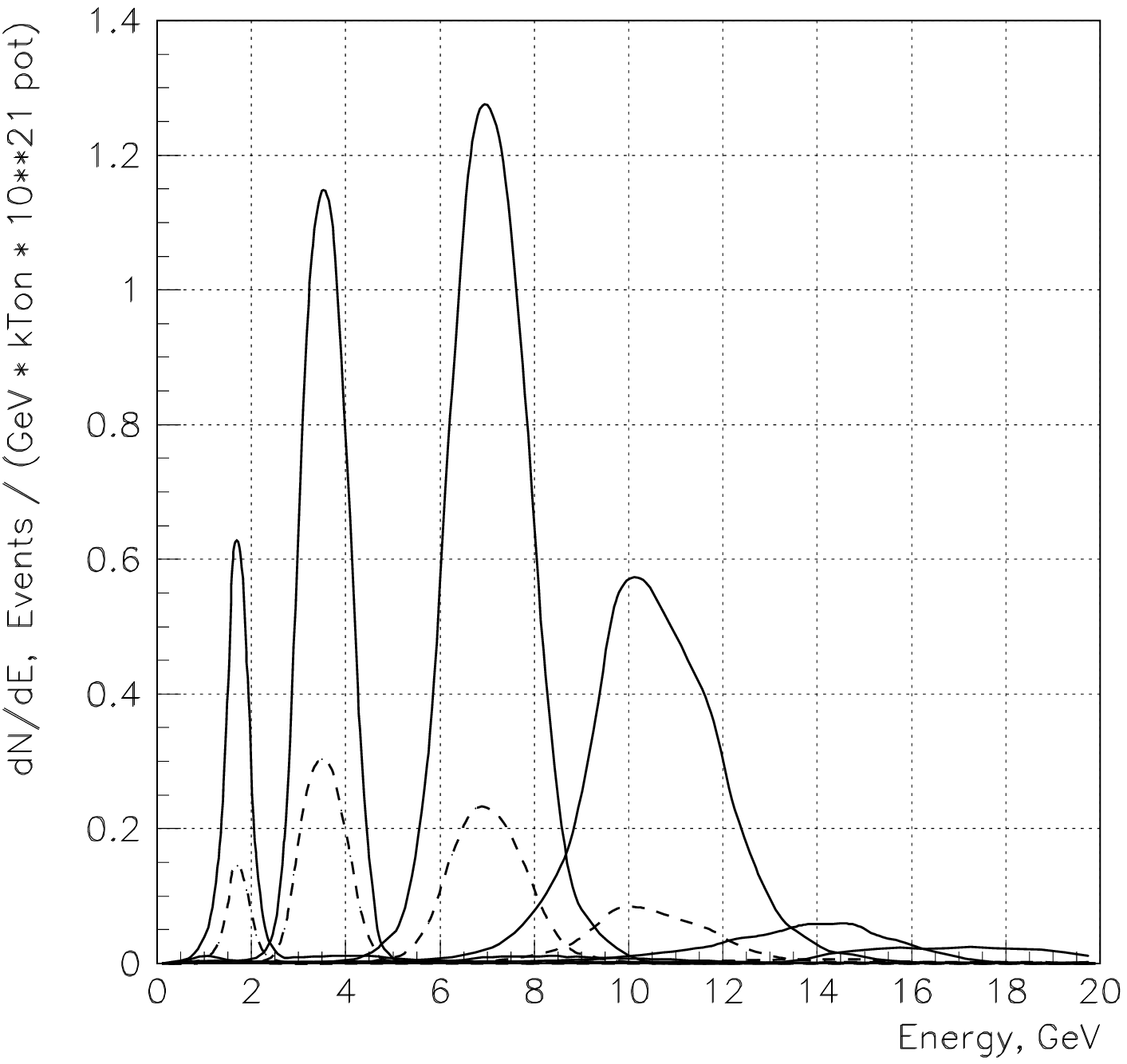,width=9.0cm}\\
\begin{minipage}{\textwidth}
{\bf Fig. 5.} Energy spectra of the $\nu_\mu$ (solid lines) and $\bar
\nu_\mu$ (dashed lines) CC events in the UNK detector for the 1.7,
3.5, 7.0, 10.5 and 14.0~GeV energy settings.
\end{minipage}
\end{tabular}
}
\vspace{0.3cm}

\begin{center}\begin{minipage}{\textwidth}
{\bf Table 3.} Focusing system and neutrino beam
parameters.  Target positions are given with respect to the
upstream end of the first horn. The baseline is $7000$~km.
\end{minipage} \\[1ex]
\begin{tabular}{|l|c|c|c|c|c|}
\hline
Beam energy range, GeV &
1.0--2.3 & 2.3--5.0 & 5.0--9.0 & 6.0--15. & 8.0--19. \\ \hline
Target Z$_{upstream}$, m & --0.34 & --0.34 & --1.94 & --2.64 & --3.34 \\ \hline
Mean energy, GeV & 1.7 & 3.5 & 7.0 & 10.5 & 14.0 \\ \hline
Energy spread HWHH, \% &
$\sim$16 & $\sim$16 & $\sim$14 & $\sim$14 & $\sim$13 \\ \hline
$\nu_\mu$ events/kTon/$10^{21}$pot &
0.45 & 1.5 & 2.8 & 2.1 & 0.33 \\ \cline{2-6}
Fraction of $\bar\nu_\mu$ events, \% &
0.8 & 0.4 & 0.3 & $\leq$0.3 & $\leq$0.3 \\ \hline
$\bar\nu_\mu$ events/kTon/$10^{21}$pot &
0.11 & 0.38 & 0.49 & 0.27 & 0.033 \\ \cline{2-6}
Fraction of $\nu_\mu$ events, \% &
$\leq$16 & $\leq$6 & $\leq$10 & $\leq$17 & $\leq$30 \\ \hline
\end{tabular}
\end{center}

\noindent
The optics layout, event rates and backgrounds are summarized in
Table~3.

%\vspace{0.5cm}
%Because the direction of a primary proton beam does not coincide
%with decay pipe axis (the angle is equal to 29 mrad), the $\nu_e$
%background comes almost entirely from the decay pipe, i.e. $\mu^+$
%and $K^+$ three-body decays give about 90\% of the $\nu_e$ CC
%events in the detector for all neutrino beam tunes.

One should note the following:
\begin{enumerate}
\item Three tune-ups of the considered focusing system  span together the
energy range from 1.7 to 7~GeV and tune-ups intermediate between these
three can be easily achieved. The more complicated
configuration of a focusing system (with obviously smaller
focusing efficiency for some particular tune-up) is required to
provide neutrino beams with $\langle E_\nu\rangle$ from 1.7 up to 13--14~GeV.
%But in any case for the 13--14~GeV neutrino beam one may
%expect the $\nu_\mu$ CC rate, which does not exceed
%0.2~events/kTon/$10^{21}$pot.
\item To obtain more intensive neutrino beams with $\sim 3$ and 
$6\;\mbox{events}/\mbox{kTon}/10^{21}$~pot for
$\langle E_\nu\rangle$ = 3.5 and 7.0~GeV, respectively, one may use a 
rectilinear focusing system with horns at the same positions but without 
dipoles. In this case the energy spreads of beams (HWHH) are of about 37\%.
\end{enumerate}
%\vspace{0.5cm}

%Further study of the focusing system including optimization of
%horns and dipoles parameters requires extensive Monte Carlo
%calculations.

\subsection{\bf  Concept of detector}
%To have enough statistics we should think about a detector
%with effective mass $ \sim 1~Mt~  level$.
%It is reasonable to use in such case the surrounding
%soil of the UNK tunnel as neutrino target. It is proposed
%to cover the UNK tunnel walls by scintillation counters.
%The length of the UNK tunnel is $\sim 21~ km$, transverse dimension ~ 5 m.
%It means that an area $\sim 10^5~ m^2$ can be used
%for a very long baseline neutrino experiment.
%Fig.? shows the transverse cross-section of the UNK tunnel
%A detector should be rather course. It is intended to use the TOF
%and energy deposition in scintillator for event identification.
The proposed baseline design for the UNK detector relies on the
known plastic scintillator extrusion technology.
The UNK detector uses extruded plastic scintillator which is
read out by wavelength-shifting (WLS) bars coupled to
photomultipliers. Some of the features which make plastic
scintillator attractive are:
\begin{itemize}
\item High efficiency for crossing particles registration;
\item  Fast scintillation timing;
\item  Ease of calibration procedure;
\item Long-term stability and reliability;
\item  Production potential - the
plastic scintillator facility at IHEP, Protvino allows to produce the full
amount of
scintillator within 2--3 years;
 \item Low maintenance --- the plastic scintillator detector
 is quite robust and will require
little maintenance in the underground tunnel conditions.
\end{itemize}
%\vspace{-2mm}
The proposed detector consists of scintillating counters (10~mm
thickness, 50~cm width and up to 6~m length) which cover the
walls of the UNK tunnel. The proposed transverse coverage of the
UNK tunnel is presented in Fig.~6.

The adjacent scintillation counters are coupled with each other
through the WLS-shifter (3~mm thickness, 10~mm width) for
light collection along the 6~m length (see Fig.~7) and thus all
counters in one 6~m cylindrical section
 form the continuous circular chain.

%The counter design shown in fig.7 is optimized to provide
%high efficiency for muon detection, good time resolution for
%background rejection and reasonable production cost.
%The detector is arranged so that muons traverse two planes
% of scintillation counters.
Scintillation counters based on polystyrene with PTP and
POPOP fluorescence dopants  can be produced by extrusion
technology at the IHEP, Protvino scintillator production facility.
%Two $25~cm$
%Scintillating strips are interlaid by a
The PMMA-based WLS-shifter bars with Kumarin-30
Fluorescence doping will be used.
The light attenuation length for such
WLS-shifter is of 3~m~\cite{Belikov,Abramov}. 
The
scintillator (decay time 2.3~ns; emission peak 420~nm;
attenuation length 20~cm at 420~nm)
and WLS bars  (decay time 2.7~ns;
emission peak 460~nm; attenuation length up to 3~m at 460~nm)
are wrapped in white material (e.g. TYVEK) and black paper.
\centerline{
\begin{tabular}{c}
\epsfig{file=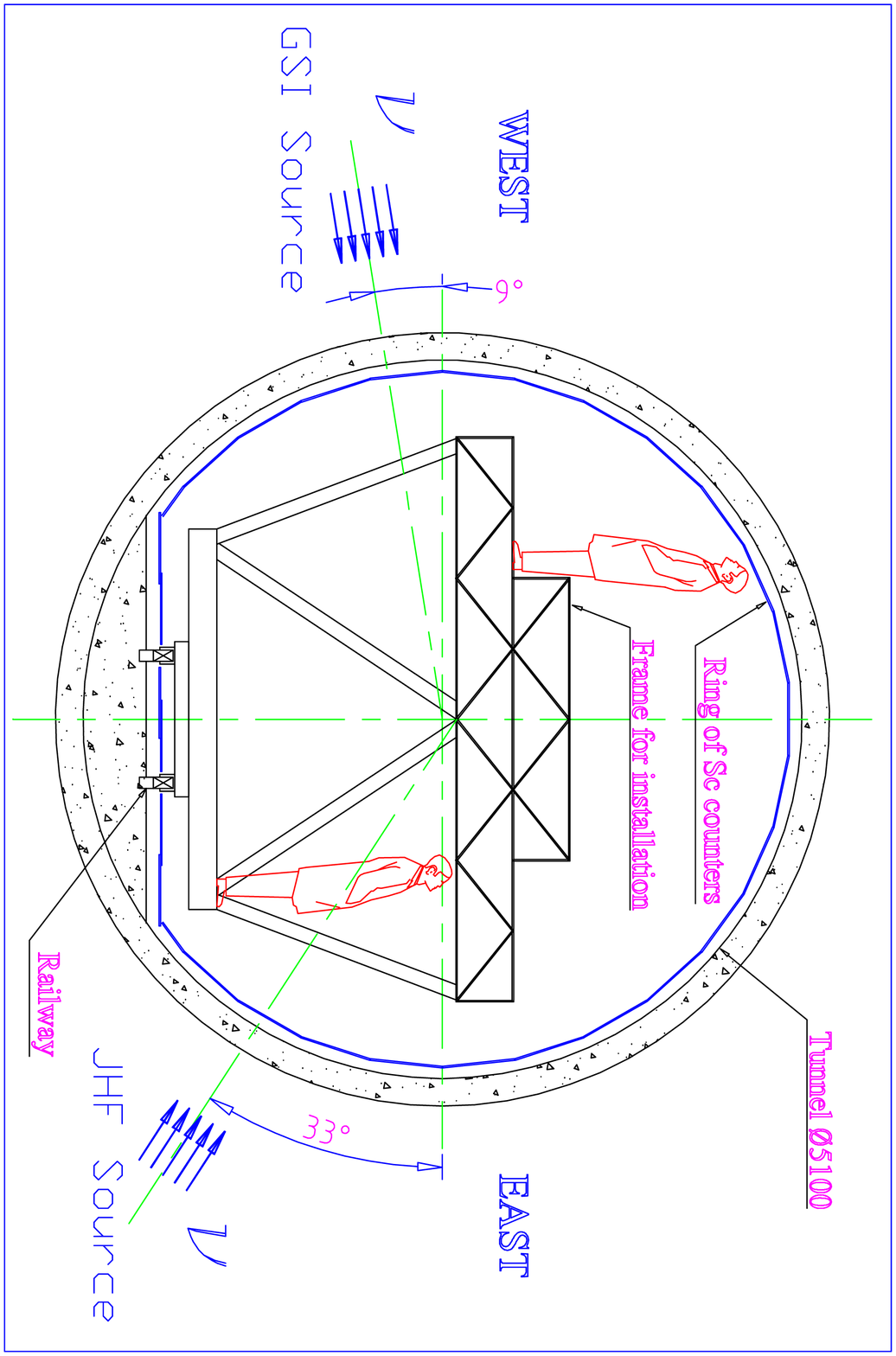,width=7.0cm,angle=90}\\
{\bf Fig. 6.} The proposed transverse coverage of the UNK tunnel.\\[3ex]
\epsfig{file=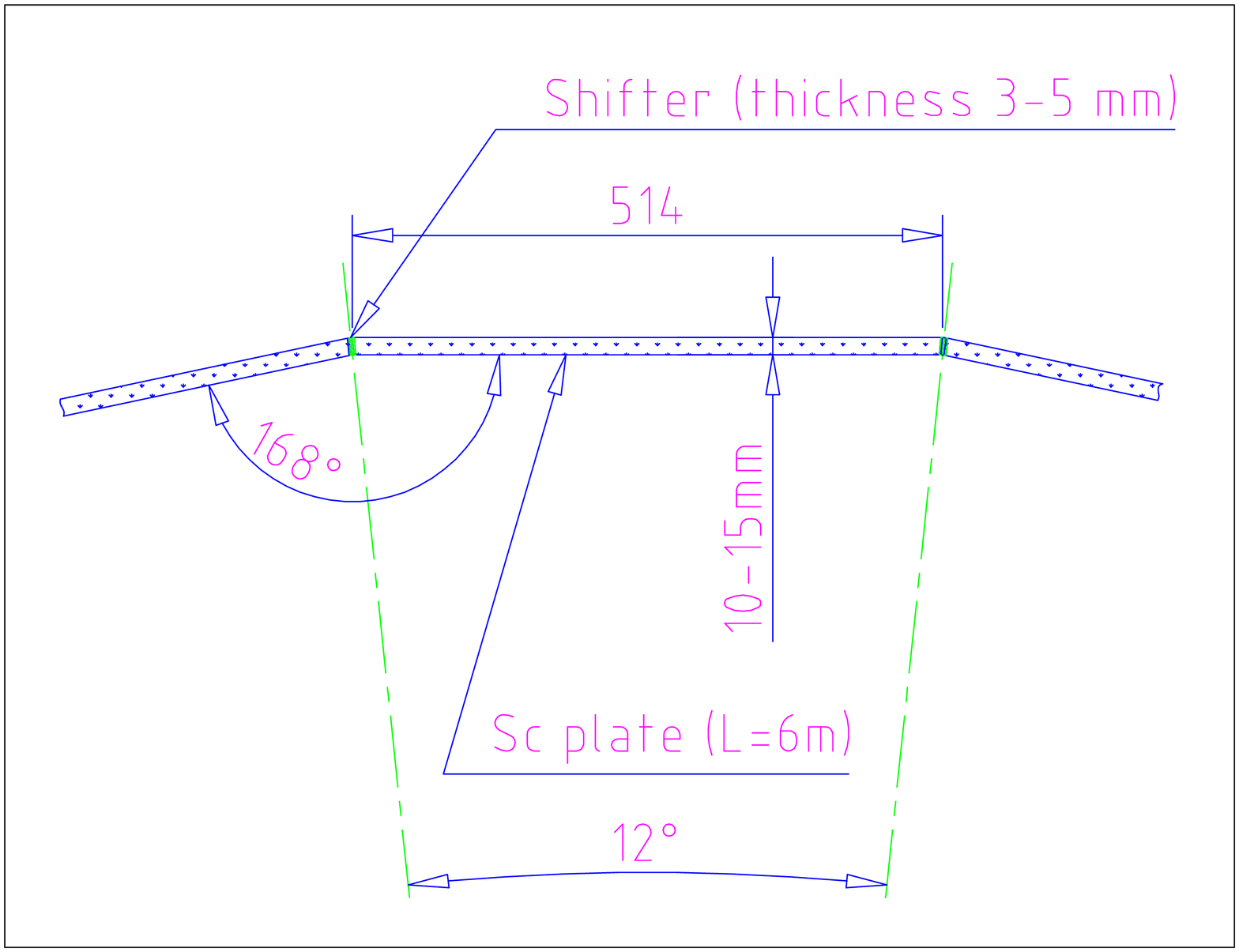,width=8.0cm,angle=90}\\
{\bf Fig. 7.} The scintillation counters coupling.
\end{tabular} }
\vspace{0.3cm}

\noindent
The light is collected from both sides of the WLS-shifter bar
 using the $1''$ PMTs. A green extended
phototube FEU-115M from MELZ (Moscow, Russia)~\cite{PHENIX} can be used
for light collection. The PMT has an average quantum efficiency of
15\%~ at 500~nm with a gain around $10^6$. The estimations show that
the signal for muon, crossing the scintillating strip
near the PMT will be $\sim 25$~ph.el.  Signals from phototubes are
delivered to the QDCs and TDCs.

The coordinate resolution is estimated as $\sim 30$~cm along the tile and 
$\sim 10$~cm in transversal direction using the TDC and QDC information.

Each cylinder section with diameter 5.1~m and length 6~m,
contains 32 counters.
%is considered as independent unit with own
%trigger for read-out.
Trigger is defined as coincidence of at least
two nonadjacent scintillation counters and performed only within
neutrino source spill time.

%Transverse coordinate for particle
%crossing the counter can get analyzing  ADC`s amplitude ratio
%values from one counter side neighbouring PMT`s. At this case
%coordinate resolution is determine by light attenuation length in
%scintillator ($20~ cm~ at~ 420~ nm$) and will be around few cm region.
%For longitudinal coordinate measurement two methods can be used -
%analyzing ADC`s amplitude ratio values and/or  TDC`s timing
%values from counter different sides PMT`s. In both cases
%coordinate resolution can estimate like $30-40~ cm$ and corresponding
%measurements is needed for real number obtaining.
% For monitoring of phototubes gain and timing a special LED based
%calibration system will be required. This system will illuminate
%each phototube with light pulses in amplitude and shape similar to
%that of a muon signal. The light will be generated by a LED pulser
%module monitored by a PIN diode and distributed to each counter
%with a light splitter by a clear 1 mm diameter plastic fiber. An
%optical connector will be mounted on each counter for the clear
%fiber. Counters timing and gain will be monitored by
% front-end-electronics
% ADCs and TDCs.

In the Table~4 main features of the UNK underground detector are given.
\begin{table}[th]
\begin{center}
{\bf Table 4.} Main parameters of the UNK detector.\\[1ex]
  \begin{tabular}{|l|c|}
  \hline
Item                                   &  Value  \\\hline
 counter dimensions            & $10\times 500\times 6000$~mm$^3$ \\\hline
time resolution                        & 1~ns  \\\hline
coordinate resolution                  & 25~cm    \\\hline
inefficiency                           & $4\cdot 10^{-5}$\\\hline
number of counters in the cylinder section & 32        \\\hline
 total number of sections             & 3500        \\\hline
total number of counters              & 112000      \\\hline
 total weight of scintillator          & 3400~Ton    \\\hline
total volume of the UNK tunnel         & 400000~m$^3$ \\\hline
  \end{tabular}
 \end{center}
\end{table}
%Each cylinder section with diameter $5.1~m$ and length $6~ m$,
%containing 32 counters, will work independently and be read out only
%in neutrino source gate using the GPS synchronization.

\section{\bf  Physics performance}
\subsection{\bf Statistics}
To estimate the statistics of the experiment we performed a Monte-Carlo
simulation of the detector response for the simple oscillation
approach of vacuum oscillations $P(\nu_{\mu}\rightarrow
~\nu_{\mu})=1-I_{\mu}\cdot \sin^2(1.27\cdot \Delta m^2\cdot L/E) $.
We used the coverage of the UNK tunnel shown in Fig.~6. We
concentrated our study on the selection of $\nu_{\mu}$ and $\bar
\nu_{\mu}$ CC events.
% For the simulation we used neutrino event
%generator developed for the SCAT BC exposed in the IHEP neutrino
%beam.
The following criteria were applied for selection:
\begin{enumerate}
\item Presence of two hits in two non-adjacent scintillation counters.
\item Time difference between hits greater than 10~ns.
\item ``Tracks'' pointing to the neutrino source.
\item Angle between neutrino and ``track'' directions less than
$30^\circ$ in each plane, and ``track'' direction below horizon.
\item $0.7<\beta=v/c<1$.
%is less than $1$.
\end{enumerate}
In the Table~5 the CC and NC event rates for different
neutrino energy settings are presented.
\begin{table}[th]
\begin{center}
\begin{minipage}{\textwidth}
{\bf Table 5.} Number of events in case of no oscillation
for $10^{7}$~s running time for each energy setting.
\end{minipage}\\[1ex]
  \begin{tabular}{|c|c|c|c|c|c|c|c|c|}
  \hline
          &\multicolumn{4}{|c|}{$\nu_{\mu}$} &
\multicolumn{4}{|c|}{$\bar \nu_{\mu}$ } \\ \cline{2-9}
          &\multicolumn{2}{|c|}{ CC } &
\multicolumn{2}{|c|}{ NC }
&\multicolumn{2}{|c|}{ CC} &
\multicolumn{2}{|c|}{NC } \\ \cline{2-9}
$E_\nu$, GeV  & GSI &JHF  &GSI &JHF &GSI  &JHF &GSI &JHF \\\hline
 1.7      & 560 &160 &  12& 4  &150  &   50 & 3 & 0.7   \\\hline
 3.5      & 3400&1100 &  70& 18 &1100  &  360 &15 & 3.3  \\\hline
 7.0      & 12000&4200&200 & 40 &2800 &  1000 &30 & 9    \\\hline
 14.0     & 2800&1000 & 34 &  9 &380  &  130  & 4 & 1.1   \\\hline
  \end{tabular}
 \end{center}
\end{table}

 As it is seen a
reasonable number of CC events can be selected for the
%"Dog Leg"
NBB
with $E_{\nu}> 2$~GeV within one year of running
($10^{7}$~s) for the JHF and GSI cases.
%For the GSI case the number of events
%should be 4 times higher due the shorter distance.

Fig.~8 shows the dependence of the detector effective mass for the
CC and NC events as a function of the neutrino energy. Fig.~9 shows
(a)~CC $\bar \nu_{\mu}$/$\nu_{\mu}$ effective mass ratio and
(b)~$\nu_{\mu}$NC$/\nu_{\mu}$CC and
 $\bar \nu_{\mu}$NC$/\bar \nu_{\mu}$CC event ratios.\\
\centerline{
\begin{tabular}{ll}
\epsfig{file=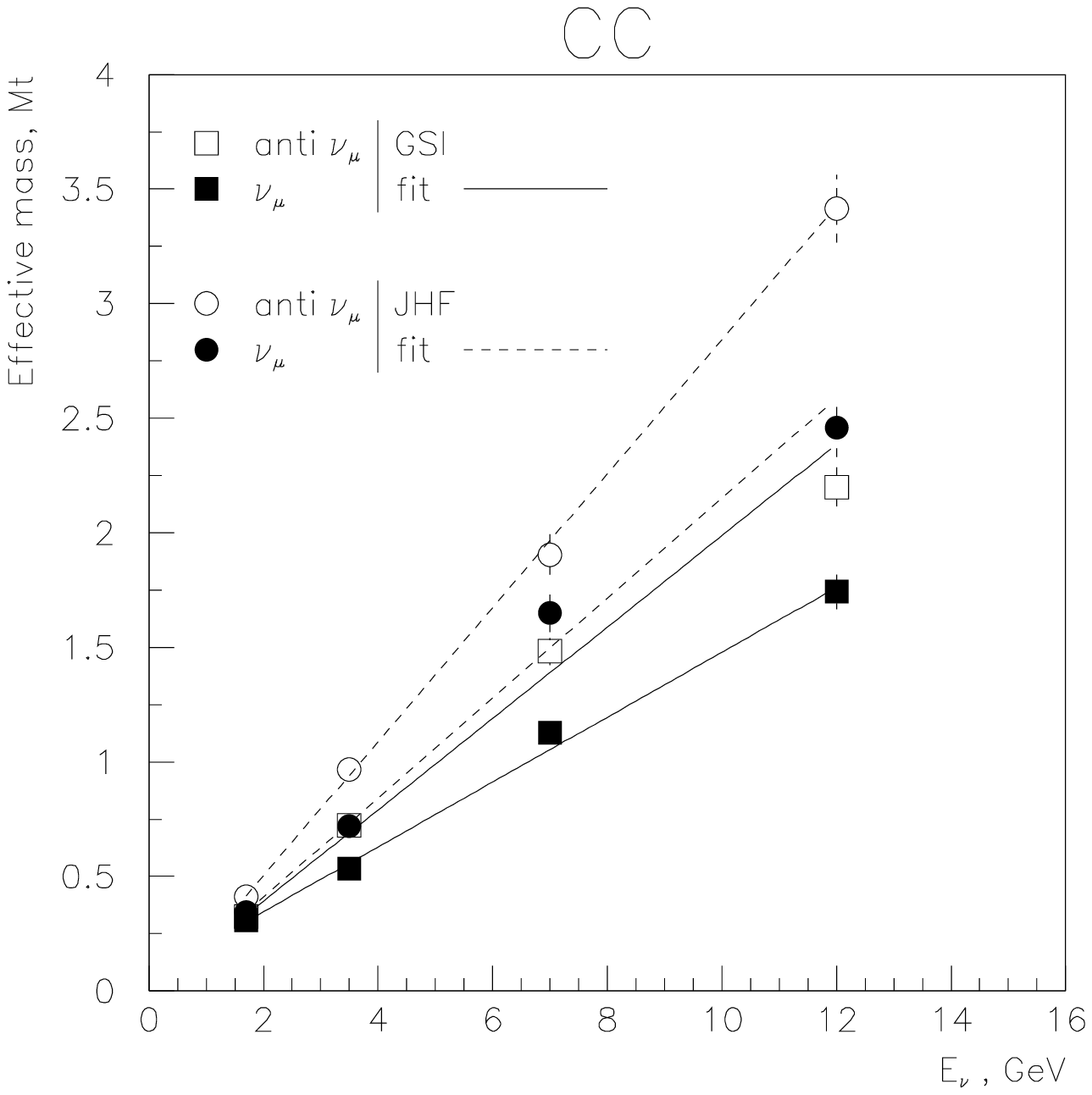,width=0.51\textwidth} &
\epsfig{file=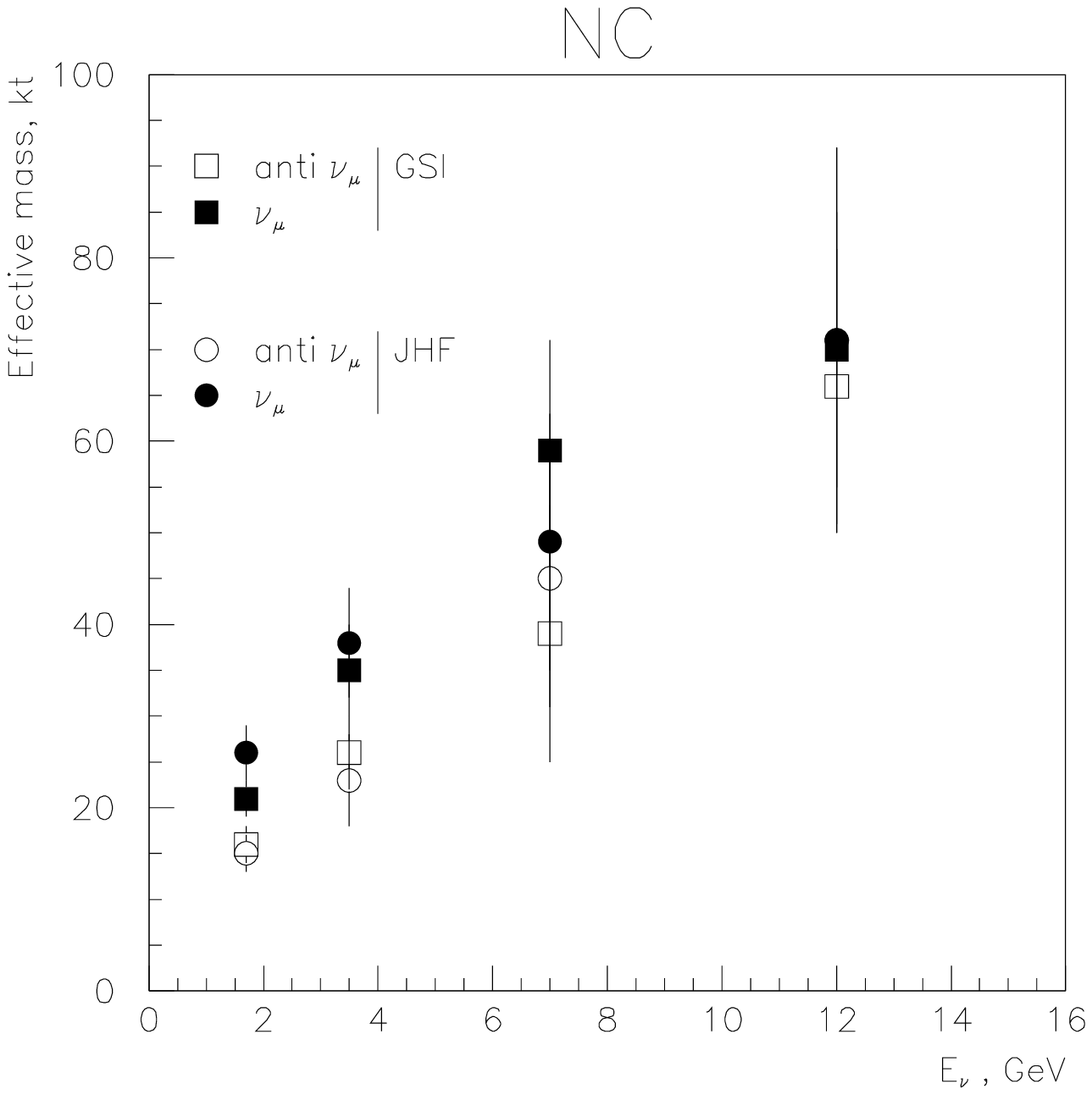,width=0.51\textwidth} \\[-4ex]
\qquad (a) & \qquad (b) \\[1ex]
\multicolumn{2}{c}{
\begin{minipage}{\textwidth}
{\bf Fig. 8.} Effective mass for (a)~CC $\nu_{\mu}$ and $\bar \nu_{\mu}$
events, (b)~NC $\nu_{\mu}$ and $\bar \nu_{\mu}$ events as a
function of the neutrino energy.
\end{minipage} }
\end{tabular} } 
\vspace{0.3cm}
 \centerline {
 \begin{tabular}{ll}
\epsfig{file=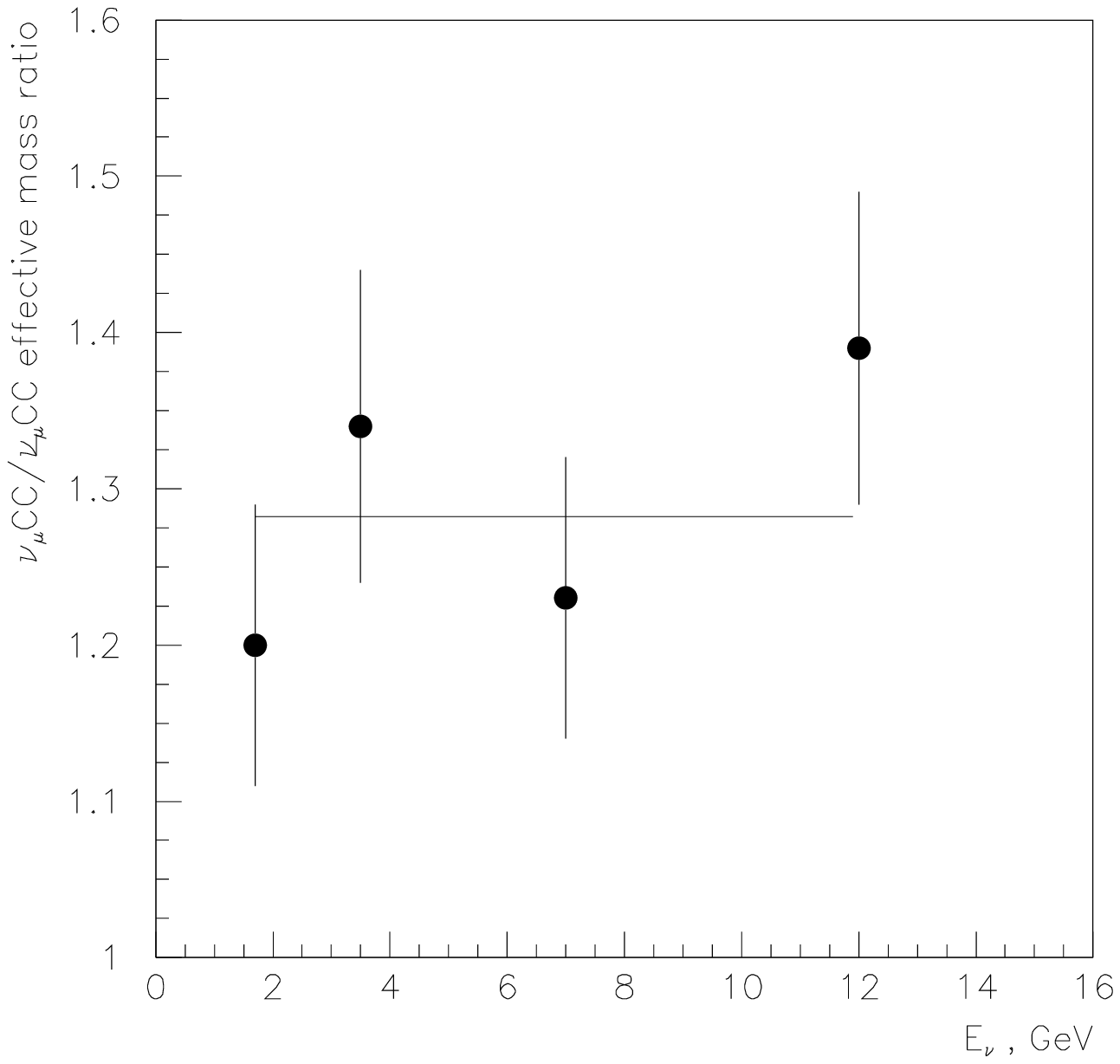,width=0.51\textwidth} &
\epsfig{file=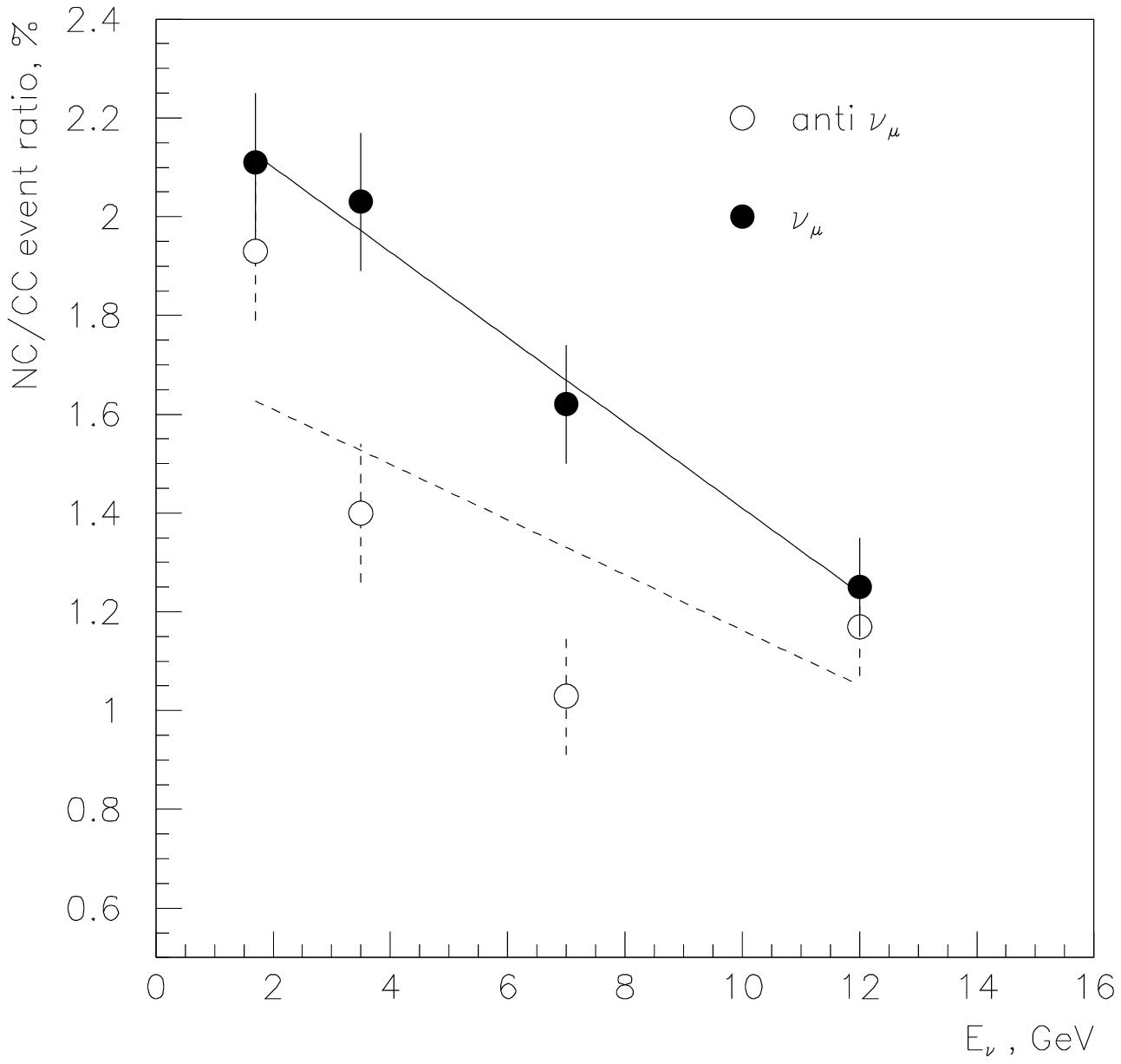,width=0.51\textwidth} \\[-4ex]
\qquad (a) & \qquad (b) \\[1ex]
\multicolumn{2}{c}{
\begin{minipage}{\textwidth}
{\bf Fig.~9.} Ratio  (a)~$\bar \nu_{\mu}$CC/$\nu_{\mu}$CC for effective
masses, (b)~$\nu_{\mu}$NC$/ \nu_{\mu}$CC and
 $\bar \nu_{\mu}$NC$/\bar \nu_{\mu}$CC for events.
 \end{minipage} }
 \end{tabular}}
  \vspace{0.3cm}

 As it is seen the CC effective mass  linearly increases
with the increase of $E_{\nu}$ reflecting the muon
range increase. Slight increase of the NC effective mass is seen as well. The
$\bar \nu_{\mu}$CC$/\nu_{\mu}$CC ratio for effective masses is
$\sim 1.3$ due to the different $y$-dependence of $\bar \nu_{\mu}$CC
and $\nu_{\mu}$CC cross sections. The NC/CC ratio is about few
 percent for  $E_{\nu}\sim 2$~GeV.
The efficiency for $\nu_e$CC and $\nu_{\tau}$CC is comparable with that
for $\nu_{\mu}$NC. Thus such kind of detector is a natural
$\nu_{\mu}$CC event selector.

%Due to the significant drop of statistics for $E_{\nu}< 2 ~GeV$
%the GSI neutrino source is suitable for the $\Delta m^2>2\cdot
%10^{-3}~eV^2$ oscillation curve pattern measurements
%and the JHF case for the $\Delta m^2<3\cdot10^{-3}~ev^2$ ones.\\

To estimate the experimental sensitivity of the oscillation
pattern measurements we select three-four neutrino energy settings
where oscillations are maximal or minimal for $\Delta
m^2=2.5\cdot10^{-3}$~eV$^2$. The statistics for each setting was
restricted to 1000 CC events assuming that there is no
oscillation. This restriction is reasonable assuming
systematic uncertainties of about $\sim 3\%$ for the expected
number of events using
the proposed NBB. In the Table~6  we present these energy
settings with the expected statistics (no oscillation), values of
pot's and exposure times for $\nu_{\mu}$ and $\bar \nu_{\mu}$
cases for the JHF and the GSI.
\begin{table}
\begin{center}
{\bf Table 6.} Estimation of running time needed
for $10^{3}$ events at each energy setting.\\[1ex]
  \begin{tabular}{|c|c|c|c|c|c|c|c|}\hline
          &     &\multicolumn{3}{|c|}{$\nu_{\mu}$} &
\multicolumn{3}{|c|}{$\bar \nu_{\mu}$ }          \\\cline{3-8}
$\nu$ source&$E_{\nu}$, GeV &Nev  &pot,$\cdot10^{21}$&time,$\cdot10^{7}$~s&
Nev  &pot,$\cdot10^{21}$ &time,$\cdot10^{7}$~s \\\hline
 JHF        & 5.7        &1000  &0.35 &0.35 &1000  &1.4  &1.4 \\\cline{2-8}
            & 7.1        &1000  &0.23 &0.23 &1000  &1.1  &1.1 \\\cline{2-8}
            & 9.4        &1000  &0.32 &0.32 &1000  &1.3  &1.3 \\\cline{2-8}
            & 14.0       &1000  &1.00 &1.00 & -    &  -  &  -  \\\cline{2-8}
            &            &Total &1.9  &1.9  &Total &3.8  &3.8 \\\hline
 GSI        & 2.0        &1000  &0.35 &1.04 & 400  &0.56 &1.67 \\\cline{2-8}
            & 4.0        &1000  &0.07 &0.22 &1000  &0.24 &0.72 \\\cline{2-8}
            & 8.0        &1000  &0.02 &0.07 &1000  &0.12 &0.36 \\\cline{2-8}
            &            &Total &0.44 &1.33 &Total &0.92 &2.75 \\\hline
  \end{tabular}
 \end{center}
 \end{table}
As it is seen from this table the oscillation pattern measurements can
be carried out in a reasonable time even in the worse
$\bar \nu_{\mu}$ case.

%{\Huge \bf BACKGROUND}\\

%Finally we estimate the sensitivity of the UNK neutrino experiment
%for determination of oscillation parameters for $\nu_{\mu}$
%disappearance. We put  values of $\Delta m^2=2.5\cdot
%10^{-3}~eV^2$ and $I_{\mu}=0.95$ and selected the following
%neutrino energy settings with correspondent $pot's$, exposure times
% and expected numbers of CC events (no oscillation):

\subsection{\bf Background}
The experiment sensitivity depends on the background level.
It is expected that the main  background is due to the cosmic muons.
Three sources of this background can be considered:
\begin{description}
\item[$\bullet\,\mu b1$] cosmic muons which coincide in direction with
accepted solid angle of muons from the neutrino source;
\item[$\bullet\,\mu b2$] cosmic muons which interact in the surrounding soil
 and produce through-going secondary particles within the accepted neutrino
source direction;
\item[$\bullet\,\mu b3$] cosmic muons which move in the direction opposite to
 the accepted solid angle for the neutrino source but with wrong TOF
identification.
\end{description}
The $\mu b1$ background can be estimated knowing the muon flux
as a function of the zenith angle $\theta $. Fig.~10 shows this
muon flux for the UNK detector for an average depth of $50~$m.
It is seen that the flux sharply drops  down to the
value of $4\cdot 10^{-7}~\mu\cdot$m$^{-2}\cdot$sr$^{-1}\cdot$s$^{-1}$ at 
$\theta =90^\circ$. This constant value for $\theta >89.5^\circ$ is defined by
muons from atmospheric neutrino interactions. It means  that we
should accept only muons below horizon
($\theta >90^\circ$). This cut 
\centerline{
\begin{tabular}{c}
\epsfig{file=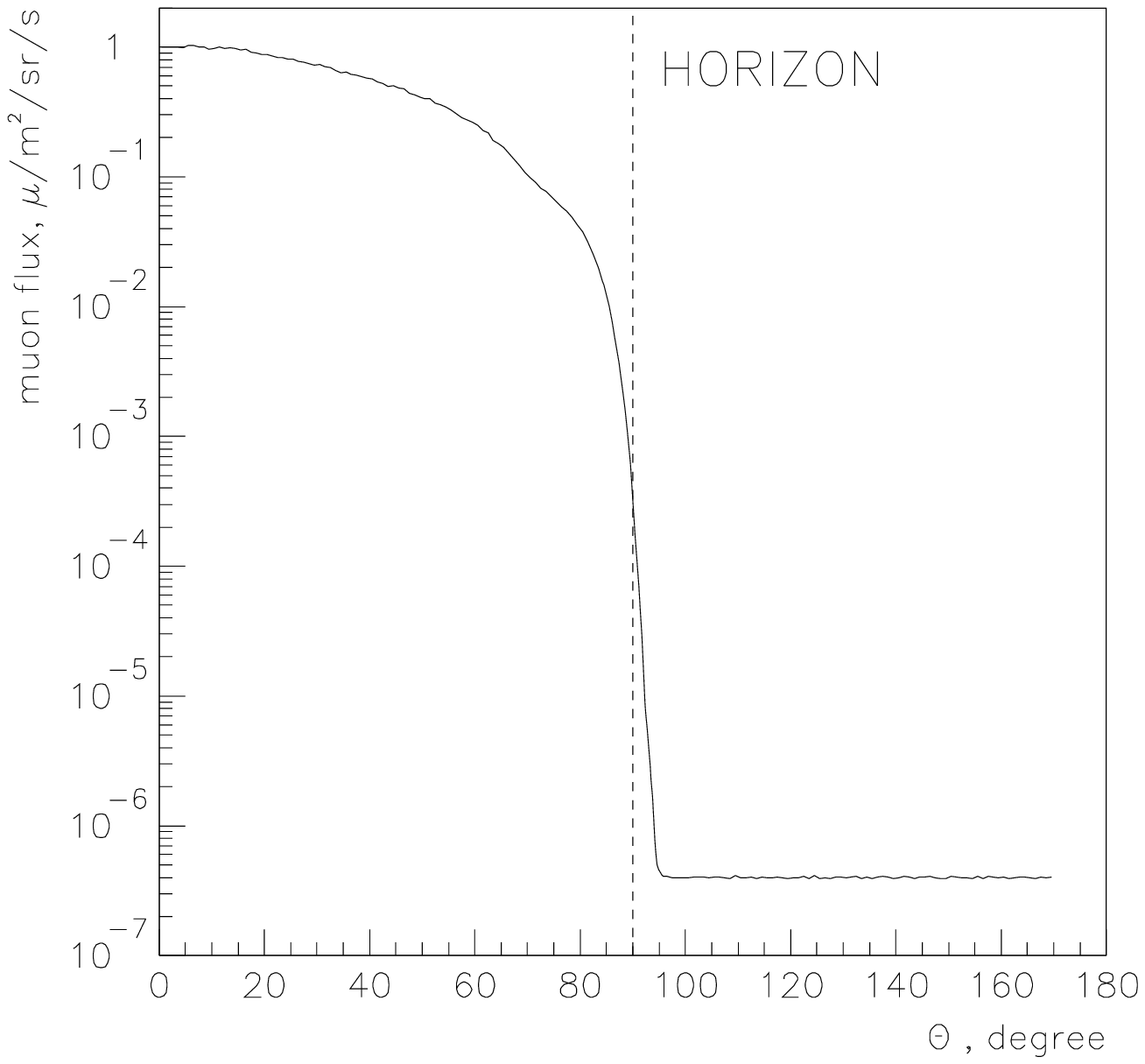,width=9.0cm}\\[-2ex]
\begin{minipage}{\textwidth}
{\bf Fig. 10.} Muon flux as a function of zenith angle $\theta $ for the
UNK underground detector (50~m depth in average).
\end{minipage}
\end{tabular}
}
\vspace{0.3cm}

\noindent
reduces the statistics for the
GSI neutrino source by $\sim 20\%$.
We estimate this background as
%\begin{center}
$N_{\mu b1}=T\cdot \Phi (\mu )\cdot \Omega \cdot S \cdot D=0.8~\mu$
%\end{center}
for an exposure time $T=10^7$~s (one year),
a solid angle $\Omega =1~sr$, a muon flux
 $\Phi (\mu )= 4\cdot 10^{-7}~\mu\cdot $m$^{-2}\cdot$sr$^{-1}\cdot$s$^{-1}$,
a detector area $S=10^5$~m$^2$ and a duty factor $D=2\cdot 10^{-6}$.

The $\mu b2$ background is estimated using the MACRO background
measurements~\cite{MACROB}. It was found that each down-going muon
produces the  $2\cdot 10^{-5}$ registered up-going particles. It
is estimated that for the UNK detector this background is $ N_{\mu
b2}=40~\mu $ for one year exposure ($T=10^7$~s). Therefore,
it is the most significant source of cosmic muon background.
It can be further reduced  using the RF structure of the beam.

We estimate the $\mu b3$ background as the value of
$ N_{\mu b3}=0.2~\mu$ for $10^7$s  assuming
$\Phi (\mu )= 1~\mu\cdot$m$^{-2}\cdot$sr$^{-1}\cdot$s$^{-1}$ and the 
$6\sigma $ separation for the TOF system ($10^{-7}$ probability for the 
Gaussian).

For the first glance the cosmic muon background is not so severe
for our experimental conditions and is $<20\%$ for the worse $\bar
\nu_{\mu}$ case at low energies. However, it should be noted that
the background is sensitive to the detector characteristics like
tails of coordinate and time resolutions which were not taken
into account. Obviously, a direct measurement of the background
in the UNK tunnel is needed.

\subsection{\bf Sensitivity }
To estimate the sensitivity of the UNK neutrino experiment
to $\nu_{\mu}$ disappearance oscillation parameters we used
the values of $\Delta m^2=2.5\cdot
10^{-3}$~eV$^2$ and $I_{\mu}=0.95$ and selected the
neutrino energy settings with the expected numbers of CC events
 (no oscillation)
presented in Table~6.
Fig.~11 shows the simulated probability values
$P=N_{osc}/N_{exp}$ with error bars,
the fitted curve and the obtained ideal oscillation curve
 for the GSI and the JHF cases as examples for the identification
and measurement of the neutrino oscillation patterns.\\

\centerline{
\begin{tabular}{ll}
\epsfig{file=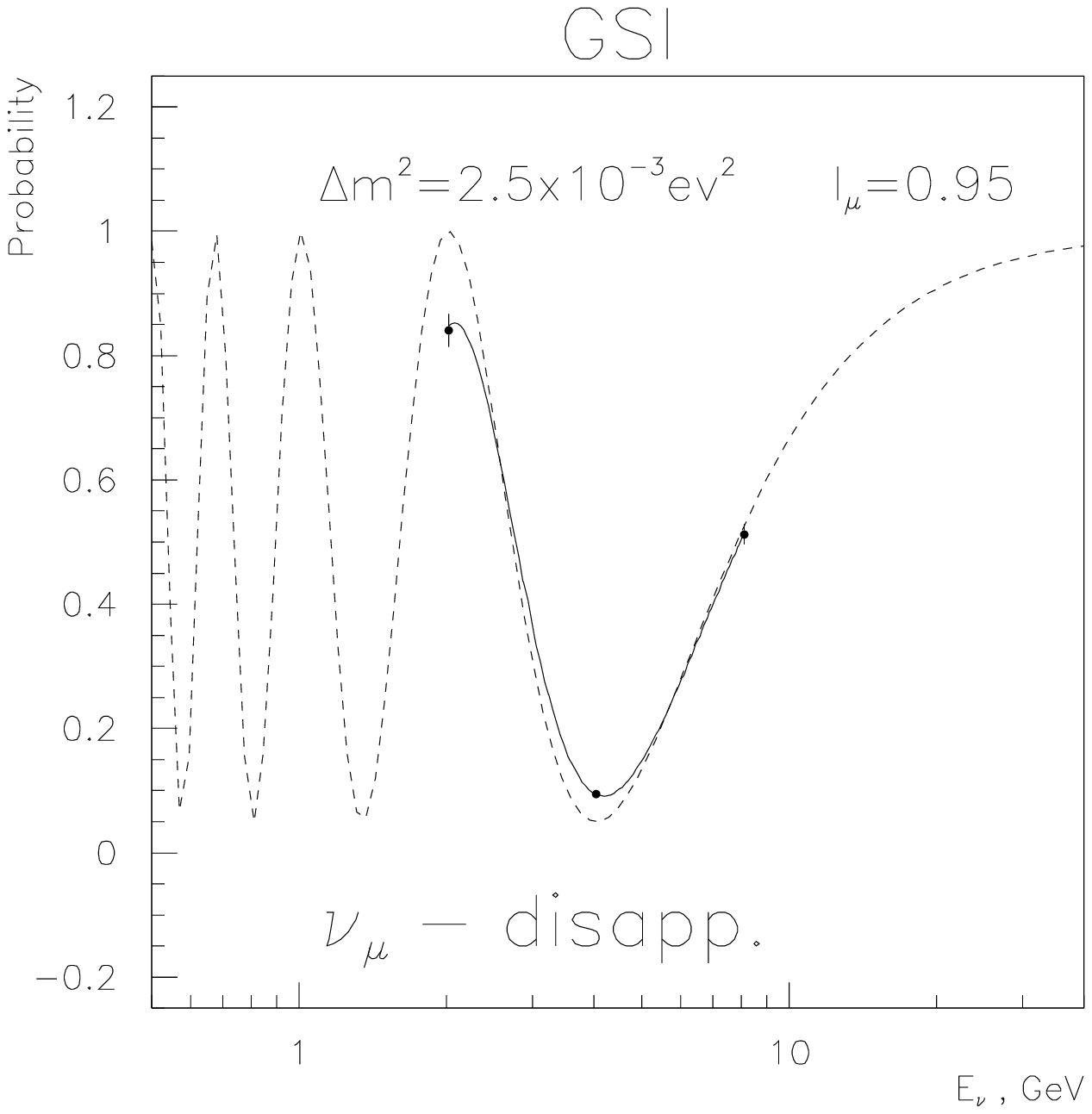,width=0.53\textwidth} &
\epsfig{file=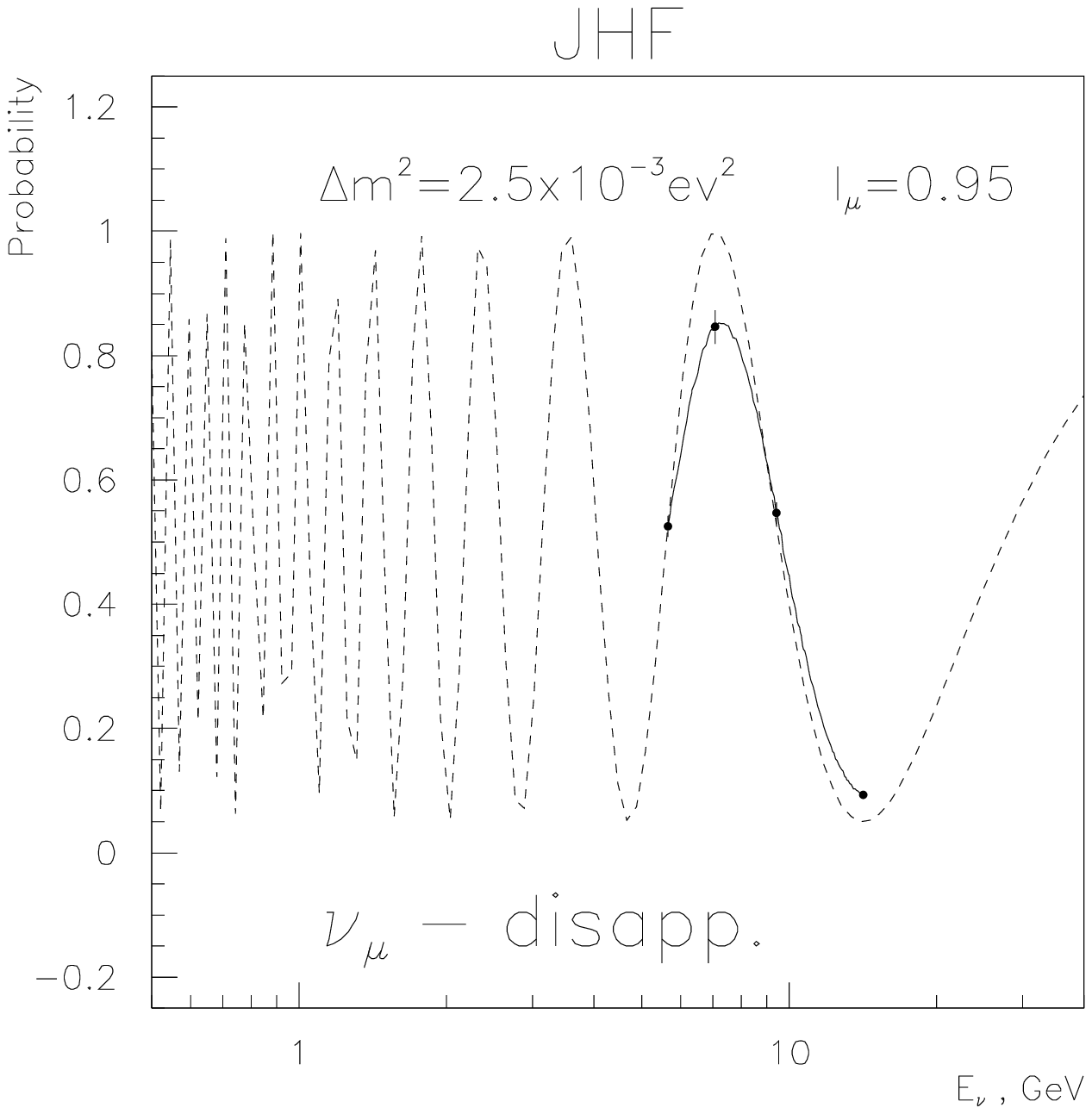,width=0.53\textwidth} \\
\multicolumn{2}{c}{
\begin{minipage}{\textwidth}
{\bf Fig. 11.} Simulated oscillation pattern (full points), fitted
(solid line) and ideal (dashed line) oscillation curves for the
GSI and the JHF cases using of $\nu_{\mu}$ data, presented
in Table~6.
\end{minipage} }
\end{tabular} }
\vspace{0.5cm}

The estimated
sensitivity for the GSI case is
\begin{eqnarray}
\sigma_{\Delta m^2} &=&2.7\cdot 10^{-5}\;\mbox{eV}^2\;\;\mbox{and}\nonumber\\
\sigma_{I_{\mu}}    &=& 0.01,\nonumber
\end{eqnarray}
and for the JHF case is
\begin {eqnarray}
\sigma_{\Delta m^2} &=&1.5\cdot 10^{-5}\;\mbox{eV}^2\;\;\mbox{and}\nonumber\\
\sigma_{I_{\mu}}   &=& 0.01.\nonumber
\end{eqnarray}
So, in this experiment the relative errors $\sim 1$\% can be achieved for $\Delta
m^2$ and $I_\mu$.

These sensitivities can be reached in the $\Delta m^2$ regions
as  presented in Table~7.

\begin{table}[th]
\begin{center}
{\bf Table 7.} The $\Delta m^2$ region accessible to the experiment.\\[1ex]
  \begin{tabular}{|l|c|}\hline
$\nu $ source &   $\Delta m^2$ in $10^{-3}$~eV$^2$         \\\hline
 GSI          &  $1.5<\Delta m^2<6$                    \\\hline
 JHF          &  $0.5<\Delta m^2<4$                 \\\hline
  \end{tabular}
 \end{center}
\end{table}

The measurement of 
$\Delta_\mu=P(\nu_\mu\to\nu_\mu)-P(\bar\nu_\mu\to\bar\nu_\mu)$
as a function of the neutrino energy can allow to
estimate the values of effective $CPT$ violation parameters 
$\delta_\mu = \Delta m^2-\Delta \bar m^2$ and
$\epsilon_\mu = I_{\mu}-I_{\bar \mu}$ with a relative precision of $\sim
2\%$ and of $\sim 1\%$, respectively.

Such a precision allows to observe oscillation or decay
neutrino patterns, to measure precisely the neutrino oscillation
parameters in the simple case  of oscillation, to search for fake and
genuine  $CPT$ violation effects at a few percent level and to search for
complicated oscillation behaviour.

% As it was pointed out in ref.\cite{CPT} the
%difference in oscillation patterns for $\nu_{\mu}$ and $\bar
%\nu_{\mu}$ disappearance cases may be an indication for the CPT
%violation in neutrino sector. Obtained sensitivity $\sigma_{\Delta
%m^2}=
%~\pm
%(2-3)\cdot 10^{-5}~eV^2$ will allow to check this prediction at
%the $10^{-4}~ev^2$ level for $\Delta m^2$ difference.

%Atmospheric muon background is important for $En < 2~ GeV$ and is
%\approxim few percent above this energy.

% It means that it is
%possible to distinguish the event direction, select MIP and
%measure tails of electromagnetic showers. It is expected that it
%is possible to select m-like and e-like events. Development of
%selection criteria is under study. For such approach statistics
%Nm-like ~ E2þfn, Ne-like,NC ~ Eþfn and therefore NC and ne event
%contribution is decreasing with increasing of neutrino energy.

\subsection{\bf Terrestrial matter effects}

The analysis given before was based on a simple approximate formula 
$P(\nu_\mu\to\nu_\mu)= 1-I_\mu\sin^2(1.27\Delta m^2\cdot L/E)$ for two-neutrino 
oscillation in vacuum, which contain two independent parameters only. 
In this section we consider more realistic approach including terrestrial 
matter effects. We restrict themselves only by analysis in the framework of 
mixing between three known neutrino. This scheme is compatible 
with the existing data on solar and atmospheric neutrino observation as well as
reactor neutrino experiments. 
Four-neutrino scheme with light sterile 
neutrino is required only by LSND experiment, which is still not confirmed. 
Moreover, recent analysis tells us that this  scheme is unsatisfactory as 
explanation of all experimental evidences for neutrino oscillations and is 
marginally acceptable~\cite{garcia}.

In the case of three-neutrino scheme, unitary matrix which transforms neutrino 
mass eigenstates into flavour ones can be written in the standard form 
\be\label{matrix}
U=\left( \begin{array}{ccc}
c_{12}c_{13}  & s_{12}c_{13} & \tilde s^*_{13}\\
-s_{12}c_{23}-c_{12}\tilde s_{13}s_{23} & 
 c_{12}c_{23}-s_{12}\tilde s_{13}s_{23} & c_{13}s_{23} \\
 s_{12}s_{23}-c_{12}\tilde s_{13}c_{23} &
-c_{12}s_{23}-s_{12}\tilde s_{13}c_{23} & c_{13}c_{23} 
\end{array} \right),
\ee
where $\tilde s_{13}=s_{13}e^{i\delta}$, $s_{ij}\equiv\sin{\theta_{ij}}$ and 
$c_{ij}\equiv\cos{\theta_{ij}}$. The 
matrix has four independent parameters: three mixing angles $\theta_{ij}$ 
($i<j$) and $CP$ violating  phase  $\delta$. In general there exist two 
additional Majorana specific phases, but they do not contribute to the lepton 
number conserving oscillations and are omitted in Eq.~\ref{matrix}.
Four parameters in matrix $U$, together
with two squared mass differences for neutrino, say $\Delta m^2_{21}\equiv 
m^2_2-m^2_1$ and $\Delta m^2_{32}\equiv m^2_3-m^2_2$,  form six parameter 
set which defines full oscillation pattern. At present, the best fit values for 
these parameters lie in two distinct statistically significant regions called 
{\it large mixing angle} (LMA) and {\it low mass}  (LOW)~\cite{garcia}. They 
are the two solutions which we will keep in mind in further investigation. We do
not intend to explore the all allowed parameter space. Rather, we are going to 
answer general questions such as the principal possibility to measure  
$CP$ violation parameter $\delta$ or the possibility to distinguish LMA and LOW
solutions.

Differential equations which describe propagation of neutrino flavour states on 
a distance $L$ look as follows
\be\label{equations}
i{d\over d\,L}\left(\begin{array}{c}
\nu_{e} \\
\nu_{\mu} \\
\nu_{\tau}\end{array}\right)
=\left[{1\over 2E}\,U
\left(\begin{array}{ccc}
-\Delta m^2_{21} & 0 & 0 \\
0               & 0 & 0 \\
0               & 0 & \Delta m^2_{32}
\end{array}\right)U^\dagger
+\left(\begin{array}{ccc}
A(L) & 0 & 0 \\
0 & 0 & 0 \\
0 & 0 & 0
\end{array}\right)\right]
\left(\begin{array}{c}
\nu_{e} \\
\nu_{\mu} \\
\nu_{\tau}\end{array}\right),
\ee
$$A(L)={1\over \sqrt{2}}G_FN_A\rho(L).$$
Here $E$ is the energy of neutrino beam, $A(L)$ is the potential induced by 
coherent interaction of electron neutrinos with electrons in the matter, $G_F$ 
is the Fermi constant, $N_A$ is Avogadro's number and $\rho(L)$ denotes the 
Earth matter density. Equations for antineutrinos are obtained 
from Eq.~\ref{equations} by replacement $(U,A)\to(U^*,-A)$. Fake $CPT$ violation
appears due to the different sign at $A$ for neutrino and antineutrino. In the 
vacuum $P(\nu_\mu\to\nu_\mu)=P(\bar\nu_\mu\to\bar\nu_\mu)$.

In our analysis we solve these equations numerically with density profile 
$\rho(L)$ taken from the Preliminary Reference Earth Model~\cite{earth}. 
As for the mixing angles and neutrino squared mass differences,  
we choose two sets of the parameters
corresponding to  the aforementioned LMA and LOW regions

\be\label{points}
\begin{array}{ccc}
 & \Delta m^2_{21}=1.4\cdot 10^{-4}\,\,\,\mbox{eV}^2 &  \\
 & \theta_{12}= 35^\circ& \,\,\,\,\,\,\mbox{Point 1 (LMA)} \\
|\Delta m^2_{32}|=2.5\cdot 10^{-3}\,\,\,\mbox{eV}^2 & & \\
\theta_{23}=40^\circ & & \\
\theta_{13}=13^\circ & & \\
& \Delta m^2_{21}=1.0\cdot 10^{-7}\,\,\,\mbox{eV}^2 & 
\,\,\,\,\,\,\mbox{Point 2 (LOW)}\\
 & \theta_{12}=35^\circ .& 
\end{array}
\ee
These values will be used as reference points in parameter space but we will 
also consider variations  in some of parameters in the experimentally
allowed limits.

To characterize the effects of fake $CPT$ violation due to the non-zero matter 
density it is convenient to introduce asymmetry between 
$\nu_\mu$ -- $\bar\nu_\mu$  oscillation probabilities
\be
A_{CPT}={P(\nu_{\mu}\to\nu_{\mu})-P(\bar\nu_\mu\to\bar\nu_\mu)\over
P(\nu_{\mu}\to\nu_{\mu})+P(\bar\nu_\mu\to\bar\nu_\mu)}.
\ee

Before discussion the results of our calculations we make some general 
remarks. \\[1ex]
1. Without loss of generality one may take for 
mixing angles and phase in matrix $U$ the following values 
$$\theta_{ij}\in [0,{\pi \over 2}]\quad\mbox{and}\quad\delta\in [-\pi,\pi].$$
From Eq.~\ref{equations} one can  show that, in the constant matter density
approximation,  $P(\nu_\mu(\bar\nu_\mu)\to\nu_\mu(\bar\nu_\mu))$ does not alter
under the change $\delta\to -\delta$.  The experiment under discussion
could not distinguish $\delta$ and $-\delta$. \\[1ex]
2.  When any
of the mixing angles in matrix $U$  is zero, the dependence
on phase $\delta$ in oscillation probabilities disappear. In the experimentally 
allowed parameter  space $\theta_{13}$ is the 
smallest angle and it is the only one which is compatible with zero 
($0\le s^2_{13}\le 0.05$ and zero value minimizes the $\chi^2$ function for 
global fit of
present data~\cite{garcia}).\\[1ex] 
%Therefore, our choice 
%is the most optimistic as for the sensitivity to the phase $\delta$.\\[1ex]
3. In the limit $\Delta m^2_{21}\to 0$ the dependence on angle $\theta_{12}$ and 
phase $\delta$ is dropped out in Eq.~\ref{equations}. Therefore, there is no 
chance to determine these two parameters in discussed 
experiments when LOW solution is realized in Nature. \\[1ex]
%LMA region is still 
%sensitive to above parameters for the reason that $\Delta m^2_{21}$ is only 
%about one order less than  $\Delta m^2_{32}$. \\[1ex]
4. In the  limit $\Delta m^2_{21}\to 0$, the changing in sign of 
$\Delta m^2_{32}$ is equivalent to passing from Eq.~\ref{equations} for neutrinos
to equations for antineutrinos, i.e.,
$P(\nu_\mu\to\nu_\mu;\,\Delta m^2_{32})
=P(\bar\nu_\mu\to\bar\nu_\mu;\,-\Delta m^2_{32})$.
Therefore, the sign of asymmetry $A_{CPT}$, if not zero, may reflect 
the sign of $\Delta m^2_{32}$ even in the case $\Delta m^2_{21}\ne 0$. \\[1ex]
5. For the LOW solution (for which $\Delta m^2_{21}=0$ is very good approximation) 
the matter influence on oscillation pattern is sensitive to the value of 
$\theta_{13}$. At $\theta_{13}=0$ Eq.~\ref{equations} decouples into two 
pieces: electron neutrino 
does not take part in oscillations while muon and tau neutrinos oscillate as in 
vacuum and $A_{CPT}$ vanishes. This is not so for the LMA solution, zero value of 
$\theta_{13}$ does not lead to the decoupling due to the non-negligible value of 
$\Delta m^2_{21}$ and deviation of $A_{CPT}$ from zero may be significant. Hence, 
the measurement of near zero
value of $A_{CPT}$ at all energies will point out on the LOW solution with small
$\theta_{13}$.

\paragraph{Numerical results} In all figures presented below we use the 
oscillation probability smeared with energy, assuming NB like beam with energy 
spread $\delta E/E=0.15$.

Figs.~12 and 13 illustrate the matter influence on the oscillation profile for 
the case of GSI and JHF, respectively. All curves correspond to phase 
$\delta=0^\circ$ and $\Delta m^2_{32}>0$. The difference between neutrino and 
antineutrino oscillation curves is seen directly, especially it is impressive for
the case of JHF. It is interesting to note, that in the last case there is not 
only difference in amplitude but also a phase shift, which reaches up to 
$\sim 1$~GeV. 

Asymmetry $A_{CPT}$ is plotted in the Figs.~14 and 15 as function of beam peak 
energy. One can see that asymmetry has absolute maximum both for GSI and JHF 
cases. As for the 
JHF case, maximum in asymmetry appears near the position of the
first maximum of the 
oscillation curve where the difference between neutrino and 
antineutrino probabilities is 
enormous. In the GSI case the absolute differences 
between neutrino and antineutrino at first minimum and maximum of oscillation 
curve are approximately the same and, therefore, maximum in asymmetry occurs at
first oscillation minimum.     
 
Thick lines in Figs.~14 and 15 corresponds to positive $\Delta m^2_{32}$, while 
thin lines to negative one. Solid lines in Figs.~14(b) and 15(b), which are for 
values of 
parameters at Point~2 (Eq.~\ref{points}), demonstrate the reflection of 
asymmetry under the changing in sign of $\Delta m^2_{32}$. Such a behaviour is 
kept, in the energy region near the asymmetry maximum, even for the parameters 
at Point~1, in spite of the not so large hierarchy between $\Delta m^2_{21}$ and 
$\Delta m^2_{32}$. Clear  distinc\-tion between the positive and negative 
signs of 
$\Delta m^2_{32}$  is lost only at lower energies. One may conclude that at 
energies near the asymmetry maximum the sign of $\Delta m^2_{32}$ is defined 
unambiguously, whatever solution, LMA or LOW, is realized in Nature.
 
To illustrate the sensitivity to the angle $\theta_{13}$ for LOW region we 
present in
Figs.~14(b) and 15(b) the additional curves for asymmetry corresponded to
$s^2_{13}=0.01$ and 0.03. It is clear that the JHF, in contrast to the GSI case, 
is in a position to resolve small values of angle $\theta_{13}$.
 
As was mentioned early, the experiment may be sensitive to the $CP$ violating
phase $\delta$ only for LMA solution. In Figs.~14(a) and 15(a) solid lines
present the asymmetry for $\delta=0$, while dashed lines are for
$\delta=180^\circ$. One can see that at low energies, say, at $\sim 1$~GeV, the 
values of asymmetry are large enough and  clearly separated for these boundary
values of $\delta$. It is in contrast with the behaviour of asymmetry for LOW
region   of parameters, where it approaches to zero at low energies. As a
consequence, the observation of non-zero value of asymmetry at low energy will
discriminate between LMA and LOW 
solutions, on the one hand, and will allow to
determine the value of $\delta$, on the other hand. The measurement of near zero
value of $A_{CPT}$ at low energies will be less informative: 
there exist some values of
parameters when oscillation patterns for LOW and LMA solutions are very close. 
Fig.~16 illustrates possible worse case  to study  for the parameters  
of  Point~1  with  $\delta=90^\circ$  and  Point~2.

One should note that the using of NB like beam is important, first of all, for 
the precise measurement of the shape of oscillation curve and, hence, for the 
precise measurement of oscillation parameters like $\Delta m^2$ or amplitude 
$I_\mu$. In contrast with this, the low energy measurement  does not require
NBB. Rather, one should use WB like beam to compensate the decreasing in 
statistics at low energies.

\bigskip
\vspace*{3cm}
\centerline{
\begin{tabular}{ll}
\qquad\qquad\qquad\qquad LMA, GSI & \qquad\qquad\qquad\qquad LOW, GSI \\
\epsfig{file=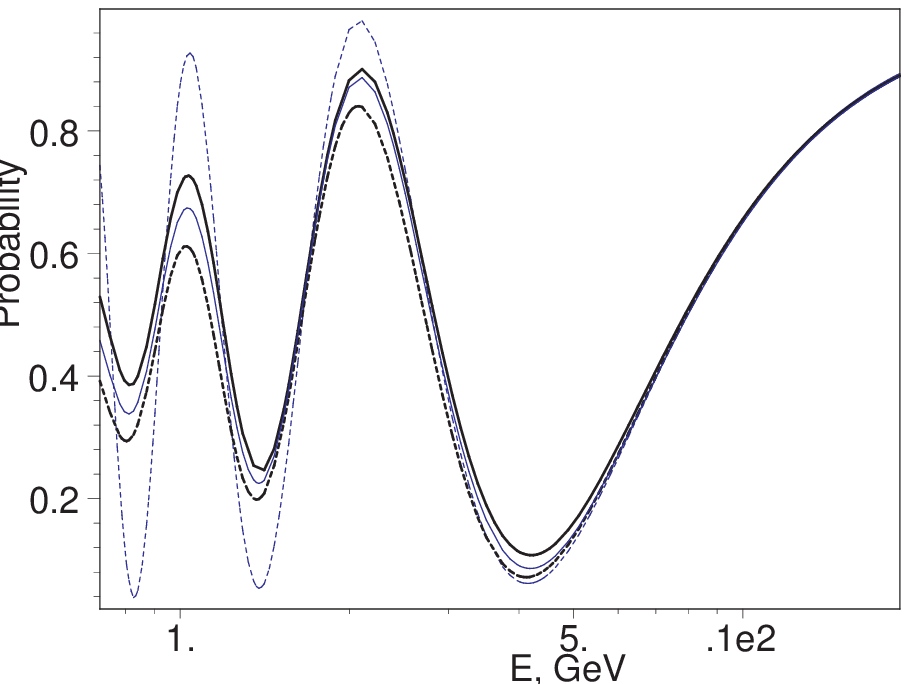,width=0.5\textwidth} & 
\epsfig{file=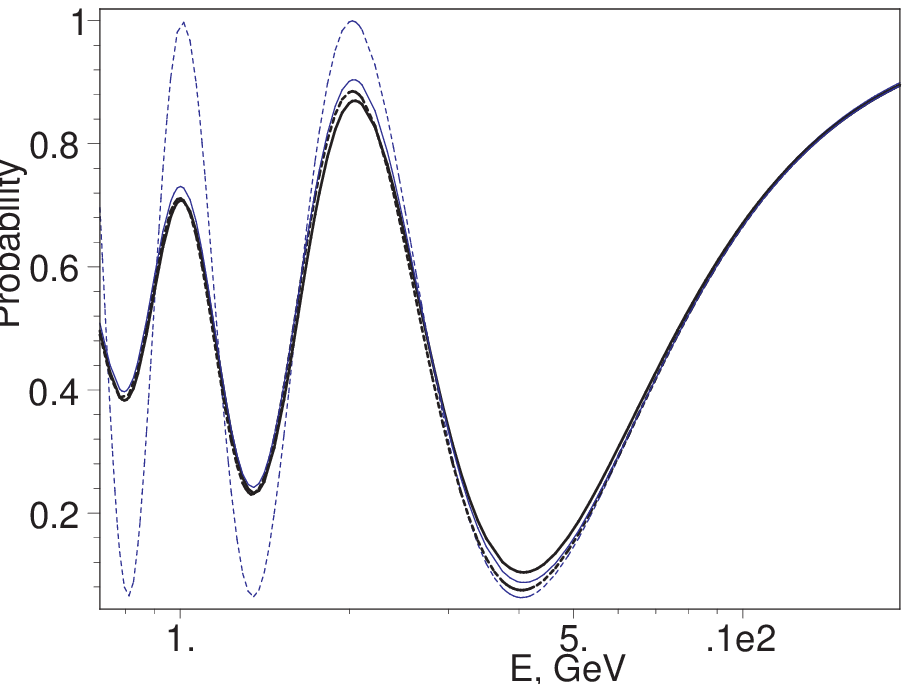,width=0.5\textwidth}\\
\qquad (a) & \qquad (b)\\[1ex]
\multicolumn{2}{c}{
\begin{minipage}{\textwidth}
{\bf Fig.~12.} Survival probability  as function of beam energy for the GSI 
($L=2000$~km): 
(a)~Point 1, (b)~Point 2 (Eq.~\ref{points}).  Thick solid line shows  
$P(\nu_\mu\to\nu_\mu)$ in matter smeared with  $\delta E_\nu/E_\nu=0.15$, thick 
dashed line shows smeared $P(\bar\nu_\mu\to\bar\nu_\mu)$ in matter. For 
comparison, smeared vacuum oscillation curve (thin solid line) and vacuum 
oscillation curve without smearing (thin dashed line) are also plotted. All 
curves correspond to $\delta=0^\circ$.
\end{minipage}
}
\end{tabular} }
\vspace{0.5cm}

\vspace{0.5cm}
\centerline{
\begin{tabular}{ll}
\qquad\qquad\qquad\qquad LMA, JHF & \qquad\qquad\qquad\qquad LOW, JHF \\
\epsfig{file=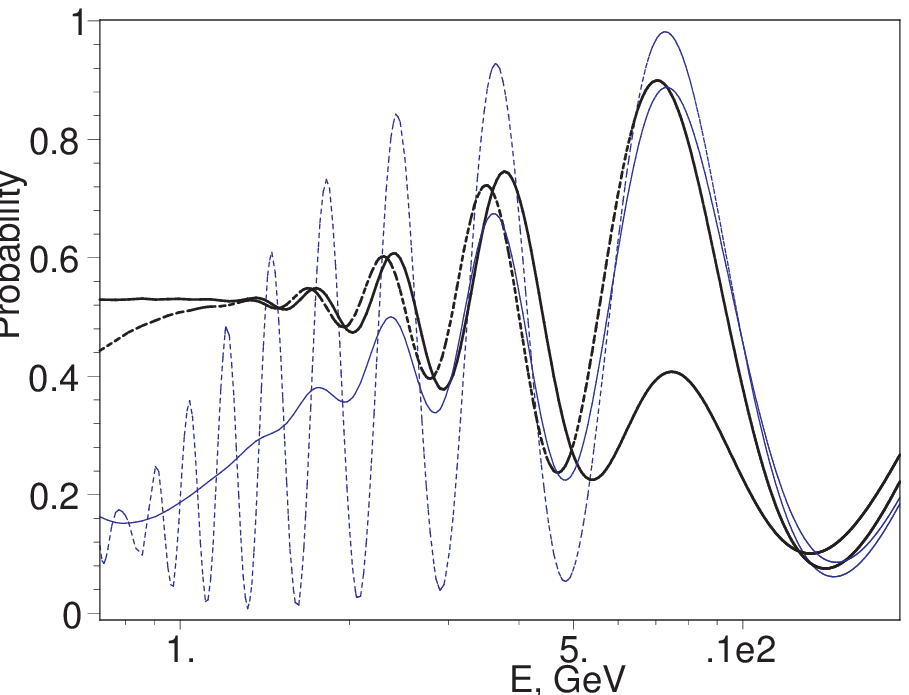,width=0.5\textwidth} & 
\epsfig{file=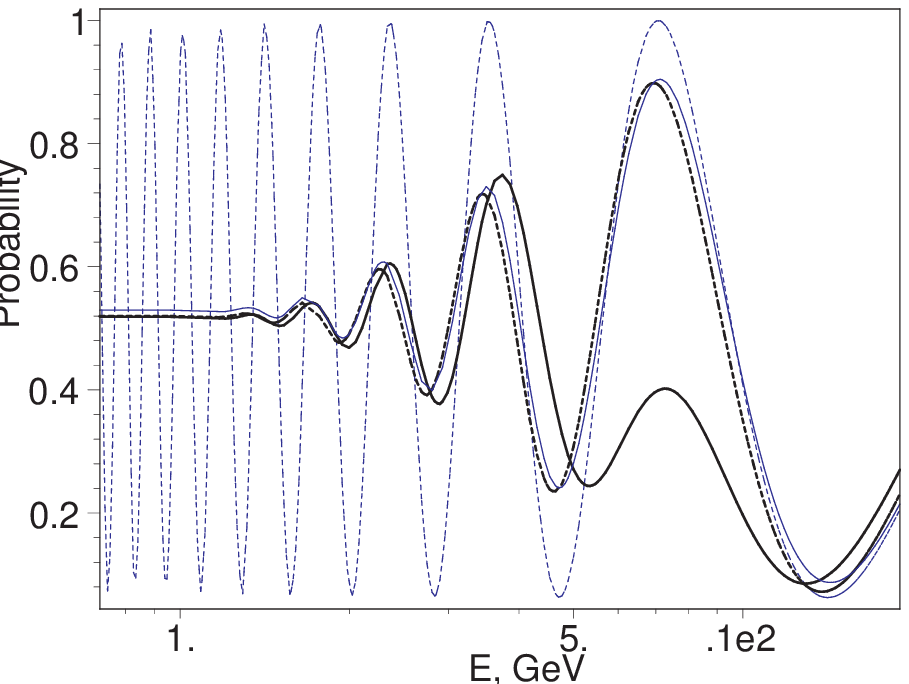,width=0.5\textwidth}\\
\qquad (a) &\qquad  (b) \\[1ex]
\multicolumn{2}{c}{
\begin{minipage}{\textwidth}
{\bf Fig.~13.} The same as in Fig. 12, but for the JHF case ($L=7000$~km). 
\end{minipage}
}
\end{tabular} }
\vspace{0.5cm}

\vspace*{3cm}
\centerline{
\begin{tabular}{ll}
\qquad\qquad\qquad\qquad LMA, GSI & \qquad\qquad\qquad\qquad LOW, GSI \\
\epsfig{file=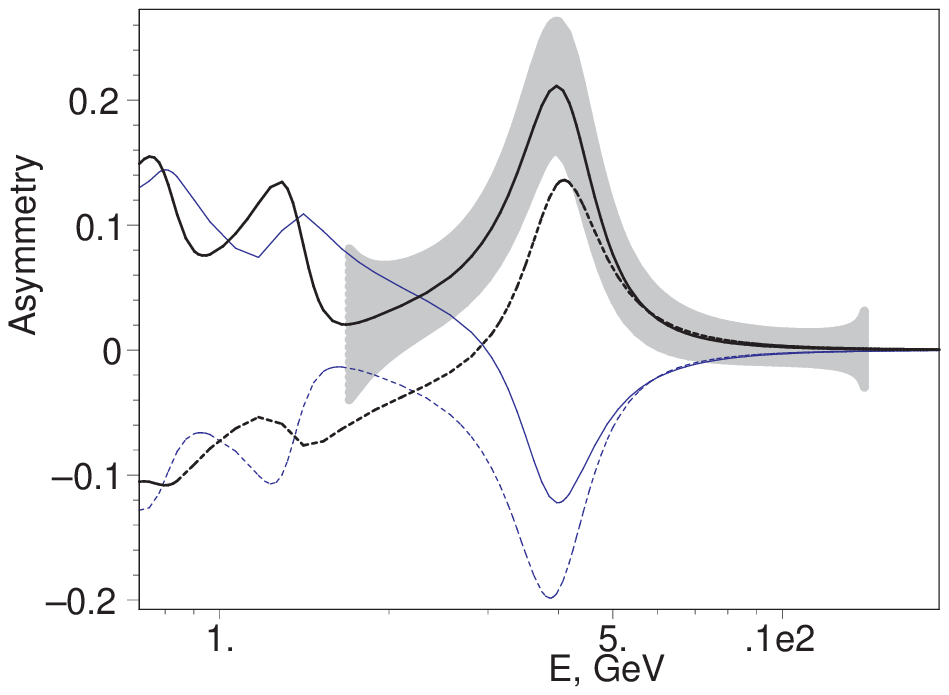,width=0.5\textwidth} & 
\epsfig{file=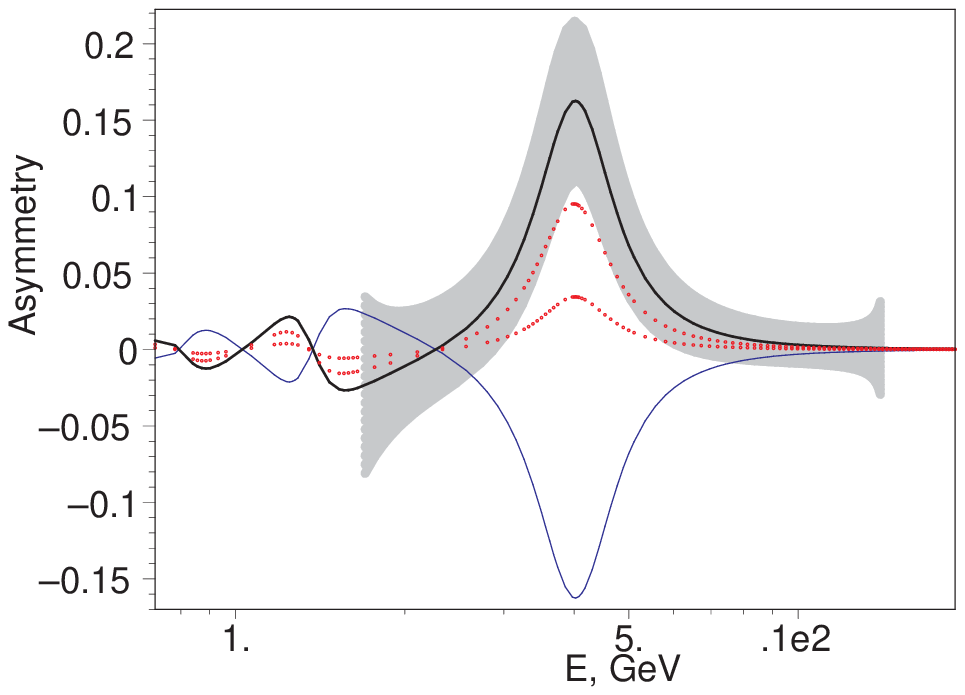,width=0.5\textwidth}\\
\qquad (a) &\qquad  (b) \\[1ex]
\multicolumn{2}{c}{
\begin{minipage}{\textwidth}
{\bf Fig.~14.} Neutrino-antineutrino asymmetry as function of beam peak energy 
for the GSI ($L=2000$~km): (a)~Point 1, (b)~Point 2 (Eq.~\ref{points}). Thick 
lines correspond to  positive $\Delta m^2_{32}$ while thin lines to negative 
ones. Solid lines are for $\delta=0^\circ$, dashed lines are for 
$\delta=180^\circ$. 
Wide grey strip shows errors in asymmetry with $\delta=0^\circ$ and 
$\Delta m^2_{32}>0$. Errors correspond to statistics for $10^7$~s running time.
Dotted lines in (b) are for $s^2_{13}=0.01$ and 0.03 ($\Delta m^2_{32}>0$);
correspondence is as follows: the smaller the angle $\theta_{13}$ the smaller the
asymmetry.
\end{minipage}
}
\end{tabular} }
\vspace{0.5cm}

\vspace{0.5cm}
\centerline{
\begin{tabular}{ll}
\qquad\qquad\qquad\qquad LMA, JHF & \qquad\qquad\qquad\qquad LOW, JHF \\
\epsfig{file=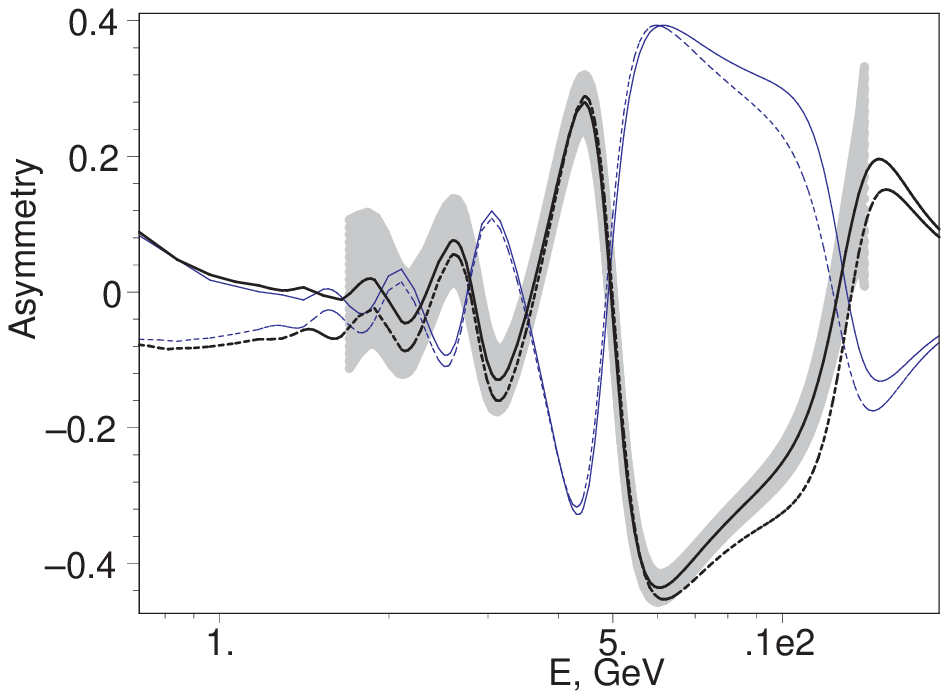,width=0.5\textwidth} & 
\epsfig{file=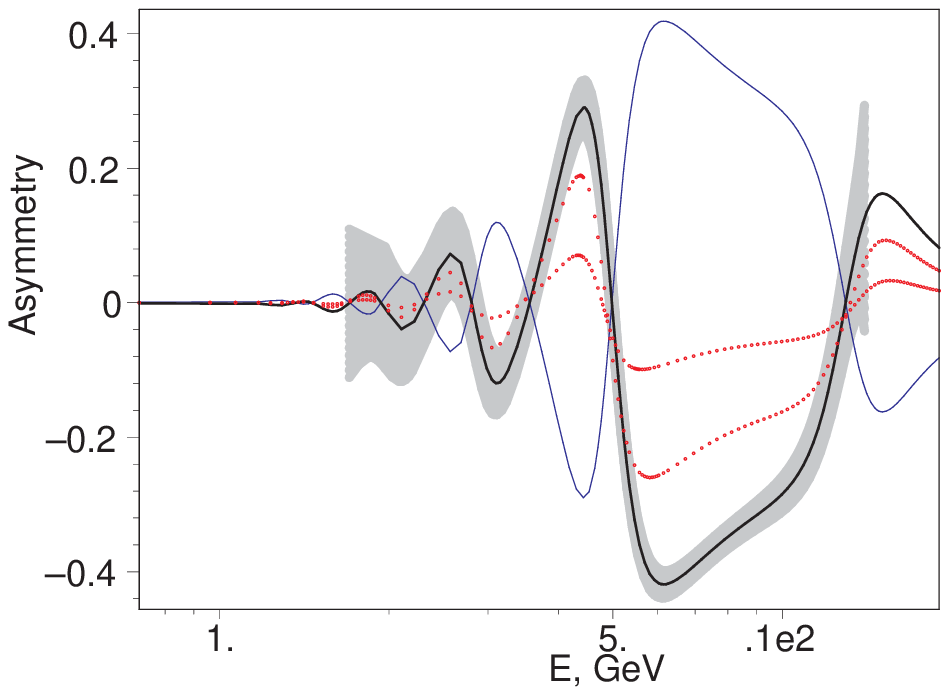,width=0.5\textwidth}\\
\qquad (a) &  \qquad (b) \\[1ex]
\multicolumn{2}{c}{
\begin{minipage}{\textwidth}
{\bf Fig.~15.} The same as in Fig. 14, but for the JHF case ($L=7000$~km). 
\end{minipage}
}
\end{tabular} }
\vspace{0.6cm}

\vspace*{2.5cm}
\centerline{
\begin{tabular}{ll}
\epsfig{file=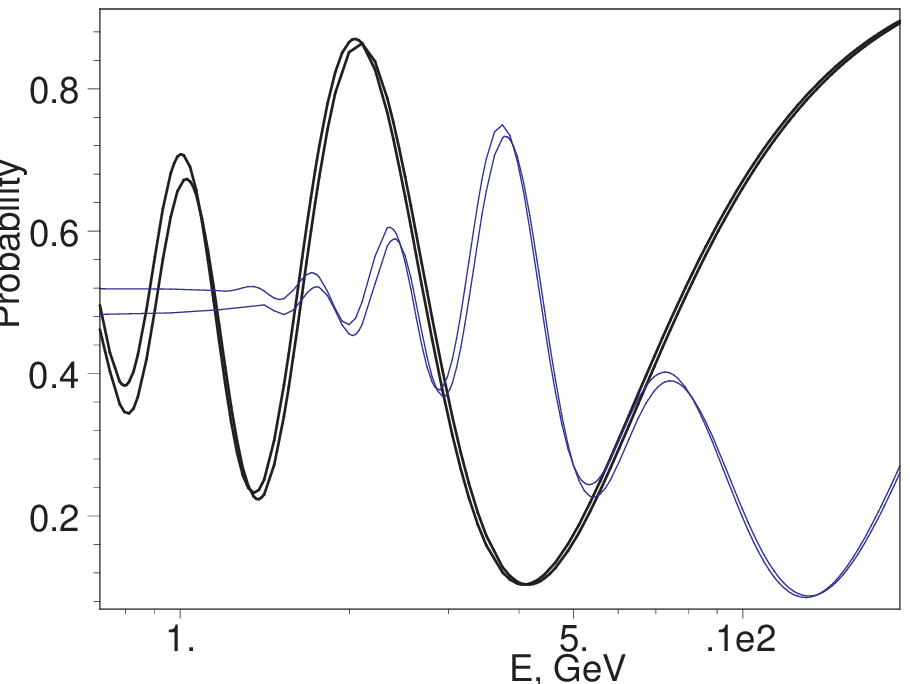,width=0.5\textwidth} & 
\epsfig{file=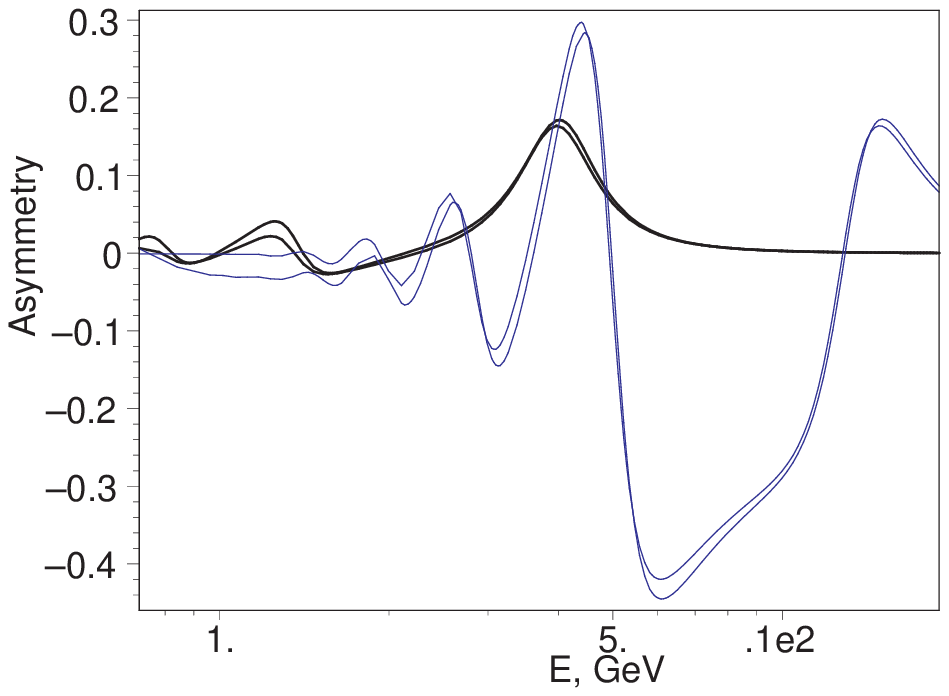,width=0.5\textwidth}\\
\qquad (a) &  \qquad (b) \\[1ex]
\multicolumn{2}{c}{
\begin{minipage}{\textwidth}
{\bf Fig.~16.} Worse case to study: Point~1 and $\delta=90^\circ$ vs. 
Point~2 and 
any  $\delta$. (a)~$P(\nu_\mu\to\nu_\mu)$; 
(b)~Asymmetry. Two thick lines correspond to GSI ($L=2000$~km), Points~1 and 2, 
two thin lines correspond to JHF ($L=7000$~km),  Points~1 and 2.  
\end{minipage}
}
\end{tabular} }
\vspace{0.5cm}

\vspace{0.3cm}
\centerline{
\begin{tabular}{ll}
\qquad\qquad\qquad\qquad LMA, GSI & \qquad\qquad\qquad\qquad LMA, JHF \\
\epsfig{file=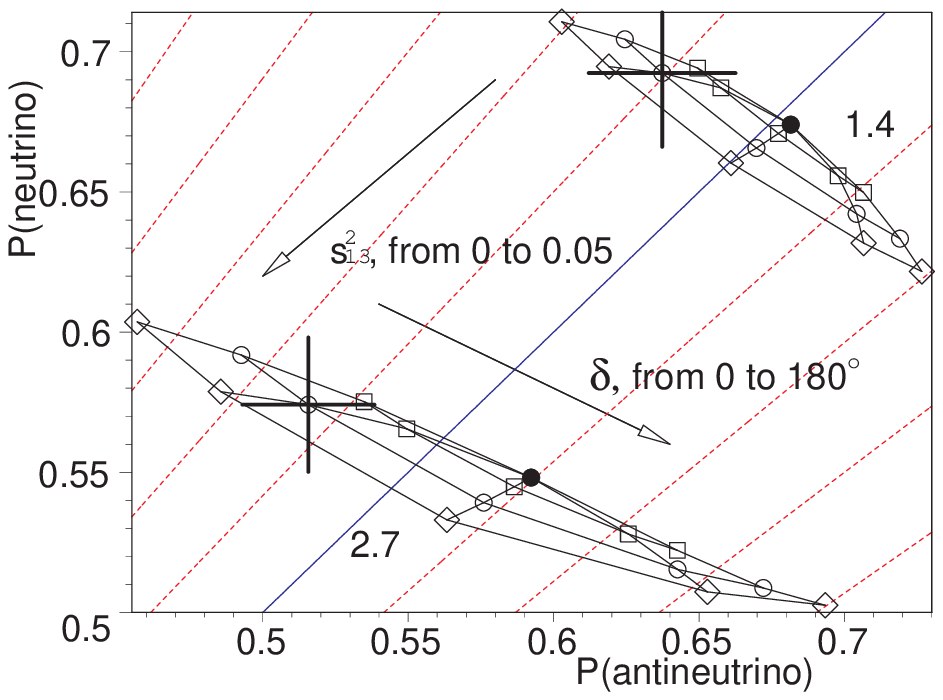,width=0.5\textwidth} & 
\epsfig{file=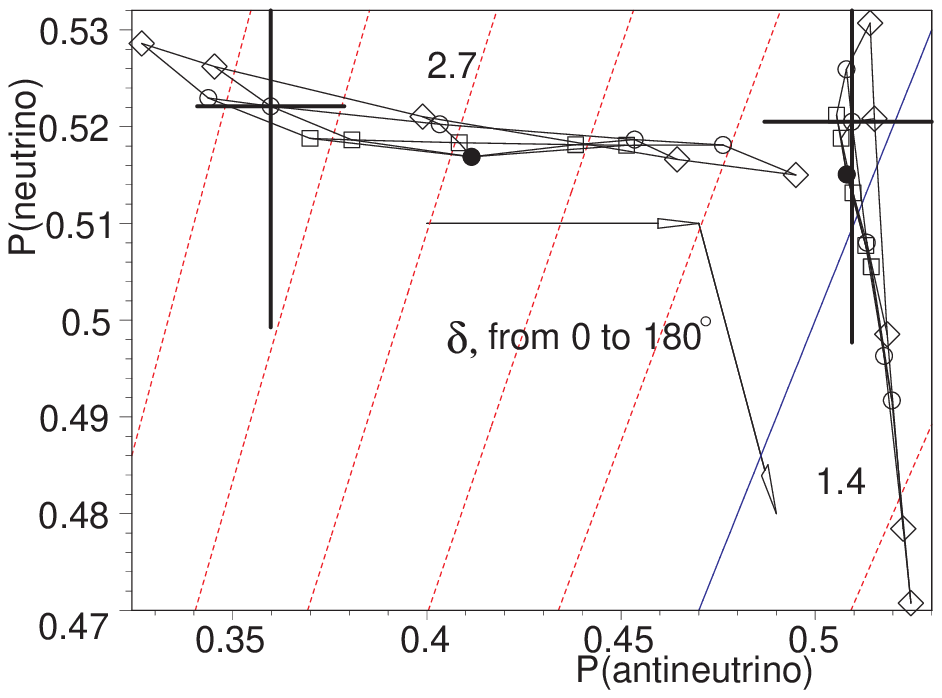,width=0.5\textwidth}\\
\qquad (a) &  \qquad (b) \\[1ex]
\multicolumn{2}{c}{
\begin{minipage}{\textwidth}
{{\bf Fig.~17.} $P(\nu_\mu\to\nu_\mu)$ vs. $P(\bar\nu_\mu\to\bar\nu_\mu)$ at
$E=1$~GeV in dependence
on phase $\delta$ and $s^2_{13}$ for Point~1: 
(a) for GSI, (b) for JHF.  Regions for $\Delta m^2_{21}=1.4\cdot 10^{-4}$~eV$^2$ 
and $\Delta m^2_{21}=2.7\cdot 10^{-4}$~eV$^2$ are marked by ``1.4'' and 
``2.7'', 
respectively.  One point corresponds 
to $s^2_{13}=0$ (black circle) and 
five points with $\delta=0^\circ$, $45^\circ$,  
$90^\circ$, $135^\circ$ and $180^\circ$ (marked by the same symbols) are plotted 
for the values of $s^2_{13}=$ 0.01 (boxes), 0.03 (circles) and 0.05 (diamonds). 
Straight line grid presents 
different values of asymmetry  
$A_{CPT}=$ 0 (solid line), $\pm 0.04$, $\pm 0.08$,~\dots (dashed lines). 
The errors
for point $s^2_{13}=0.03$ and $\delta=45^\circ$, which correspond to the 
statistics of $10^3$ events in the absence of oscillations, are plotted.}  
\end{minipage}
}
\end{tabular} }
\vspace{0.5cm}

The sensitivity to $\delta$ depends on the value of
$s^2_{13}$ and the maximal sensitivity is for maximal allowed value of
$s^2_{13}=0.05$. To comprehend the influence of $s^2_{13}$ we calculated
$P(\nu_\mu\to\nu_\mu)$ and $P(\bar\nu_\mu\to\bar\nu_\mu)$ at  beam energy 
$E=1$~GeV for various values of $s^2_{13}$ and $\delta$. Fig.~17 
presents the results of these calculations. 
It contains two distinct regions which correspond to $\Delta
m^2_{21}=1.4\cdot 10^{-4}$~eV$^2$ and $\Delta m^2_{21}=2.7\cdot 10^{-4}$~eV$^2$.
One can see that for the case of 
GSI there is one-to-one
correspondence between points in the plane 
($P(\nu_\mu),P(\bar\nu_\mu)$) and values ($s^2_{13},\delta$). The typical errors,
plotted for one point ($s^2_{13}=0.03, \delta=45^\circ$), allows to conclude that
parameters $s^2_{13}$ and $\delta$ could be, in principle, disentangled and
determined simultaneously. The situation for the case of JHF is more involved. 
The one-to-one correspondence is lost and the relative variations in oscillation
probabilities with $s^2_{13}$ and $\delta$ are smaller than in the GSI case.
Qualitatively this can be understood as the increasing role of matter effects. 
The last ones become dominant even at relatively low energies, where the 
sensitivity  to $\delta$ is maximal.

Considered in this section topics allow to make a comparison between GSI and 
JHF as the sites for neutrino beam source. If LMA solution is realized in Nature,
GSI can provide more information on oscillation parameters in the low 
energy measurements. JHF is more sensitive to the parameters of LOW region at 
higher energies where the matter effects maximize asymmetry $A_{CPT}$.

\section{\bf  Conclusion}
The VLBL neutrino oscillation experiment
 (baseline of 2000--7000~km) with the UNK $\sim 1$~MTon
underground detector will provide:
\begin{itemize}
\item Observation of the oscillation patterns for the
 $\nu_{\mu}$ and $\bar \nu_{\mu}$  disappearance in the
 $0.5\cdot 10^{-3}<\Delta m^2<6\cdot
 10^{-3}$~eV$^2$ range using the NB neutrino beam.
%of $\Delta m^2$ and $I_{\mu}$ with a sensitivity
%\begin {eqnarray}
%{\sigma_{\Delta m^2}\over \Delta m^2 }& \sim &1~\% \mbox{~~~and}\nonumber\\
%{\sigma_{I_{\mu}}\over I_\mu} & \sim & 1~\% \nonumber
%\end{eqnarray}
%both for $\nu_{\mu}$ and $\bar\nu_{\mu}$ cases.
\item Direct observation of the matter effects by comparison of $\nu_\mu$ and 
$\bar\nu_\mu$ oscillation curves.
\end{itemize}
It allows to extract important physical information such as 
\begin{enumerate}
\item Measurement of the $\Delta m^2_{32}$ and the intensity of oscillations with
the accuracy at the level of $\sim 1$\%.
\item Measurement  of $\sin^2\theta_{13}$ down to the value of $\sim 0.01$.
\item Unambiguous determination of the sign of $\Delta m^2_{32}$ by the sign 
of asymmetry $A_{CPT}$ for $\sin^2\theta_{13}\gtrsim 0.01$.
\item Distinction between LMA and LOW solutions outside the case of some 
exceptional values of mixing matrix parameters.
\item Determination of $CP$ violating phase $\delta$ in the case of LMA solution.
\end{enumerate}
Thus, the experiment allows to prove the standard scenario of 
neutrino oscillations, to measure precisely essential oscillation parameters as 
well as to search for new phenomena.

\subsection*{\bf  Acknowledgements}

The authors are thank\-ful to A.~Lo\-gu\-nov, A.~Miag\-kov, R.~Nahn\-hauer, 
V.~Rya\-bov, V.~Tsa\-rev and N.~Tyu\-rin for useful discussions. 
One of us (VA) is
very grateful to the Organizers of the 1-st International Workshop
on Nuclear and Particle Physics at 50-GeV PS (KEK, Tsukuba, Japan,
Dec.10--12, 2001) for giving him opportunity to present the talk
on this matter at the Workshop.

%2. At the point of full nm disappearance it is possible to check
%nm " nt and nm " ns hypothesis using NC events, and to establish
%more precise limit sin2 2q13 < 0.02   (today sin2 2q13 < 0.1)
%using e-like events.

%\end{document}

\end{document}